\documentclass{revtex4}
\usepackage{amssymb}
\usepackage{amsfonts}
\usepackage{amsmath}
\usepackage{graphicx,comment}
\usepackage{float}   
 \usepackage{xcolor}

\begin{document}

\title{Isochronous $n$-dimensional nonlinear PDM-oscillators:
linearizability, invariance and exact solvability}
\author{Omar Mustafa}
\email{omar.mustafa@emu.edu.tr}
\affiliation{Department of Physics, Eastern Mediterranean University, G. Magusa, north
Cyprus, Mersin 10 - Turkey,\\
Tel.: +90 392 6301378; fax: +90 3692 365 1604.}

\begin{abstract}
\textbf{Abstract:}\ Within the standard Lagrangian settings (i.e., the
difference between kinetic and potential energies), we discuss and report
isochronicity, linearizability and exact solubility of some $n$-dimensional
nonlinear position-dependent mass (PDM) oscillators. In the process,
negative the gradient of the PDM-potential force field is shown to be no
longer related to the time derivative of the canonical momentum, $\mathbf{p}%
=m\left( r\right) \mathbf{\dot{r}}$, but it is rather related to the time
derivative of the pseudo-momentum, $\mathbf{\pi }\left( r\right) =\sqrt{%
m\left( r\right) }\mathbf{\dot{r}}$ (i.e., Noether momentum). Moreover,
using some point transformation recipe, we show that the linearizability of
the $n$-dimensional nonlinear PDM-oscillators is only possible for $n=1$ but
not for $n\geq 2$. The Euler-Lagrange invariance falls short/incomplete for $n\geq 2$ under PDM settings. Alternative invariances are sought, therefore.
Such invariances, like \emph{Newtonian invariance} of Mustafa \cite{42}, effectively authorize the use of the exact solutions of one system to find the solutions
of the other. A sample of isochronous $n$-dimensional nonlinear
PDM-oscillators examples are reported.


\end{abstract}

\author{}
\maketitle

\section{Introduction}

A standard textbook Lagrangian is given by the difference between kinetic
and potential energies, otherwise it is classified as a non-standard one. \
Such standard presentation renders the total energy of a dynamical system to
be an integral of motion (i.e., a constant of motion) and consequently a
conserved quantity. The kinetic $T=\frac{1}{2}m_{\circ }\,\dot{x}^{2}$ and
the potential energies $V(x)$ are two correlated quantities, therefore. That
is, if in the standard constant mass $m_{\circ }$ harmonic oscillator
Lagrangian%
\begin{equation}
L\left( x,\dot{x};t\right) =\frac{1}{2}m_{\circ }\,\dot{x}^{2}-\frac{1}{2}%
m_{\circ }\,\omega ^{2}x^{2}\text{ };\text{ \ }\dot{x}=\frac{dx}{dt},
\end{equation}%
for example, the coordinate $x$ is transformed/deformed is such a way that $%
x\rightarrow \sqrt{Q\left( u\right) }u$, then the velocity $\dot{x}$ would
transform/deform in a completely different manner so that $\dot{x}%
\rightarrow $ $\sqrt{m\left( u\right) }\dot{u}$. In this case, the
Lagrangian in the new coordinate system reads%
\begin{equation}
L\left( u,\dot{u};t\right) =\frac{1}{2}m_{\circ }\,m\left( u\right) \dot{u}%
^{2}-\frac{1}{2}m_{\circ }\,\omega ^{2}Q\left( u\right) u^{2}.
\end{equation}%
Where, the dynamical equations for the two systems, (1) and (2), are governed by the textbook Euler-Lagrange invariance under coordinate transformation (or any other physically feasible alternative invariance). That is,%
\begin{equation}
\frac{d}{dt}\left( \frac{\partial }{\partial \dot{x}}L\left( x,\dot{x}%
;t\right) \right) -\frac{\partial }{\partial x}L\left( x,\dot{x};t\right) =0=%
\frac{d}{dt}\left( \frac{\partial }{\partial \dot{u}}L\left( u,\dot{u}%
;t\right) \right) -\frac{\partial }{\partial u}L\left( u,\dot{u};t\right). 
\end{equation}%
Yet, the straightforward relation between the dimensionless scalar functions $Q\left(
u\right) $ and $m\left( u\right) $,%
\begin{equation}
\sqrt{m\left( u\right) }=\sqrt{Q\left( u\right) }\left( 1+\frac{Q^{\prime
}\left( u\right) }{2Q\left( u\right) }u\right) ;\,Q^{\prime }\left( u\right)
=\frac{dQ\left( u\right) }{du},
\end{equation}%
identifies the correlation between the kinetic and potential energies of the so called position-dependent mass (PDM) harmonic oscillators. This correlation, moreover, clarifies how each energy term would adapt to any change in the other. Metaphorically speaking, if a PDM-particle, $M\left( u\right)
=m_{\circ }\,m\left( u\right) $, is moving with a velocity $\dot{u}$, then its PDM kinetic energy $T=$ $\frac{1}{2}m_{\circ }\,m\left( u\right) \dot{u}%
^{2}$ is correlated with a PDM potential force field $V\left( u\right) =%
\frac{1}{2}m_{\circ }\,\omega ^{2}Q\left( u\right) u^{2}$. Only under such
PDM assumptions that the Lagrangian dynamics exactly reproduces the
Newtonian ones and adheres to the textbook Hamilton"s least action principle.
Yet, only under such coordinate transformation that the total energy is
conserved and is an integral of motion. Moreover, it is obvious that this
relation suggests that $m\left( u\right) \neq Q\left( u\right) $, and $%
m\left( u\right) =Q\left( u\right) $ is valid if and only if the mass is constant (i.e., retrieving the usual textbook constant mass settings).
Therefore, the PDM-harmonic oscillator's Lagrangian and Hamiltonian should
be written as%
\begin{equation}
L\left( x,\dot{x};t\right) =\frac{1}{2}m_{\circ }\,m\left( x\right) \dot{x}%
^{2}-\frac{1}{2}m_{\circ }\,\omega ^{2}Q\left( x\right)
x^{2}\Longleftrightarrow H\left( x,p;t\right) =\frac{p^2}{2m_{\circ
}\,m\left( x\right)} +\frac{1}{2}m_{\circ }\,\omega ^{2}Q\left(
x\right) x^{2},
\end{equation}%
Otherwise, the system would not only lose the total energy conservation but
also it would violate Hamilton's least action principle. Throughout the current study, we
shall use $m_{\circ }\,=1$ unless otherwise mentioned.

On the other hand, among the most prominent PDM oscillators are the
Mathews-Lakshmanan oscillators \cite{1} described by the nonlinear dynamical
equations%
\begin{equation}
\ddot{x}\mp\frac{\lambda x}{1\pm\lambda x^{2}}\dot{x}^{2}+\frac{\omega ^{2}}{%
1\pm\lambda x^{2}}x=0\text{ };\text{  }\ddot{x}=\frac{d^{2}x}{dt^{2}},
\end{equation}%
that admit simple harmonic oscillators solutions of the form%
\begin{equation}
x=A\cos \left( \Omega t+\varphi \right) \text{ };\text{ }\Omega ^{2}=\frac{%
\omega ^{2}}{1\pm\lambda A^{2}}.
\end{equation}%
Obviously, the frequencies are restricted by the conditions that $\Omega
^{2}=\omega ^{2}/\left( 1\pm\lambda A^{2}\right) $ and are amplitude dependent
ones. Consequently, the nonlinear oscillators lose their isochronicity. This
is mainly attributed to the non-standard nature of the PDM Lagrangian%
\begin{equation}
L\left( x,\dot{x};t\right) =\frac{1}{2}\left( \frac{\dot{x}^{2}-\omega
^{2}x^{2}}{1\pm\lambda x^{2}}\right) 
\end{equation}%
adopted to work out the dynamical equations of motion in (6). It is clear that the PDM potential force field is chosen to be $V\left( x\right) =\omega
^{2}x^{2}/\left[ 2\left( 1\pm\lambda x^{2}\right) \right] $ which, in turn, renders the total energy non-conserved for the reasons discussed above. This issue is discussed in more details in section II. Nevertheless, such non-standard Lagrangians structure have inspired a great research interest in PDM settings, both in classical and quantum mechanics (c.f., e.g., the sample of references \cite%
{1,2,3,4,5,6,7,8,9,10,11,12,13,14,15,16,17,18,19,20,21,22,23,24,25,26,27,28,29,30,31,32,33,34,35,36,37,38,39,40,41,42,43,44,45,46,47}%
). 

In fact, the nonlinear differential form of the PDM Euler-Lagrange equations of (6) represents some peculiar special cases of the quadratic (i.e., with an $\dot{x}^{2}$ term) Li\'{e}nard-type nonlinear differential equation%
\begin{equation}
\ddot{x}+F\left( x\right) \dot{x}^{2}+G\left( x\right) =0\text{.}
\end{equation}%
Which is a very interesting equation because of its immense applicability and usefulness both in physics and mathematics \cite%
{1,2,3,4,5,6,7,8,9,10,11}. The linearizability and isochronicity of which have invited a vast number of interesting research studies in many fields (c.f., e.g., \cite{35,36,37,38,39,40,41,42,43,44,45,46}). Tiwari et al. \cite%
{2} and Lakshmanan and Chandrasekar \cite{3}, for example, have used Lie point symmetries and asserted that in the case of eight parameter symmetry group, the one-dimensional quadratic Li\'{e}nard type equation (9) is linearizable and isochronic. It should be mentioned, nevertheless, that the Mathews-Lakshmanan oscillators (6) are linearizable via some nonlocal point transformations \cite{3,4,5}, though rendered non-isochronuos.

In this work, however, we shall be interested in the generalization of such
nonlinear PDM-oscillators for any physically admissible and viable
PDM-settings. Therefore, we focus our attention on the class of standard PDM
Lagrangians/Hamiltonians (5) and the linearizability of their dynamical
equations. Their isochronicity (i.e., with amplitude-independent
frequencies), moreover, shall be sought and preserved in the process (c.f.,
e.g., \cite{35,36,37,38,39,40,41}). Hereby, isochronous $n$-dimensional
nonlinear PDM-oscillators form the subject of the current methodical proposal. Consequently, we organize our paper in such a way that the methodology of our proposal is made clear and comprehensive to serve for
viable/feasible pedagogical implementations of isochronous nonlinear
PDM-oscillators.

In section II, we discuss and analyze the Mathews-Lakshmanan (ML) nonlinear PDM-oscillators (6) and  recollect some preliminaries (within their non-standard Lagrangians/Hamiltonians presentations) so that their generalization to any PDM $m\left( x\right) $ settings is made feasible and safe. Following the standard Lagrangian/Hamiltonian setting (1) to (5), we obtain the PDM potential force field for the ML-PDM, $m\left( x\right) =1/(1\pm\lambda x^{2})$, and report the corresponding isochronuos PDM-nonlinear oscillators. We discuss and report, in section III, the correlation between negative the $n$-dimensional gradient of the PDM potential force field (i.e., the $n$-dimensional PDM force vector) and the pseudo-momentum $\mathbf{\pi }\left( r\right) =\sqrt{m\left( r\right) }%
\mathbf{\dot{r}}$ \cite{5,7} (i.e., Noether momentum \cite{6}). We show that negative the gradient of the PDM potential force field is no longer the time derivative of the canonical momentum, $\mathbf{p}=m\left( r\right) \mathbf{%
\dot{r}}$, but it is rather related with the time derivative of the pseudo-momentum, $\mathbf{\pi }\left( r\right) =\sqrt{m\left( r\right) }%
\mathbf{\dot{r}}$ (as in (35) below). Moreover, we discuss Euler-Lagrange (EL) invariance ( in $n$-dimensions)  between constant-mass and PDM systems. We show that the EL-invariance falls short for $n\geq 2$  and hence we report on two alternative ways to secure invariance between the two systems for the $n$-dimensional case. In section IV we introduce our isochronuos  $n$-dimensional PDM oscillators and report on their linearizability and invariance. Illustrative examples are given in section V.  We give our concluding remarks in section VI.

\section{Mathews-Lakshmanan nonlinear oscillators and their standard
Lagrangians structures}

In the generalization of the non-standard Mathews-Lakshmanan oscillators
Lagrangian (8) to cover PDM settings, one should keep in mind that
Lagrangian (8) is rewritten as%
\begin{equation}
L=\frac{1}{2}m\left( x\right) \dot{x}^{2}-\frac{1}{2}m\left( x\right) \omega
^{2}x^{2}\,.
\end{equation}%
This would imply the Euler-Lagrange dynamical system%
\begin{equation}
\ddot{x}+\frac{m^{\prime }\left( x\right) }{2m\left( x\right) }\dot{x}%
^{2}+\left( 1+\frac{m^{\prime }\left( x\right) }{2m\left( x\right) }x\right)
\omega ^{2}x=0.
\end{equation}%
Obviously, only under the assumption that%
\begin{equation}
\left( 1+\frac{m^{\prime }\left( x\right) }{2m\left( x\right) }x\right) =%
\text{ }m\left( x\right) ,
\end{equation}%
would the PDM function read%
\begin{equation}
m\left( x\right) =\frac{1}{1+\lambda x^{2}}.
\end{equation}%
Which is indeed the PDM used in the Mathews-Lakshmanan oscillator (6).
However, a question of delicate nature arises here as to "what would be the
standard form of the potential force field for the Mathews-Lakshmanan (ML)
PDM particle in (13)?  The answer to this question lies in rewriting
Lagrangian (10) as that of (5) and use the Euler-Lagrange invariance result
in (4). Namely, we use%
\begin{equation}
L=\frac{1}{2}m\left( x\right) \dot{x}^{2}-\frac{1}{2}\omega ^{2}Q\left(
x\right) x^{2}\text{ };
\end{equation}%
to imply the dynamical equation of motion%
\begin{equation}
\,\ddot{x}+\frac{m^{\prime }\left( x\right) }{2m\left( x\right) }\dot{x}%
^{2}+\omega ^{2}\frac{Q\left( x\right) }{m\left( x\right) }\,\left( 1+\frac{%
Q^{\prime }\left( x\right) }{2Q\left( x\right) }x\right) x=0.
\end{equation}%
Which is linearizable into a simple harmonic oscillator%
\begin{equation}
\ddot{q}+\omega ^{2}q=0\text{ };\text{ }q=A\cos \left( \omega t+\varphi
\right) ,
\end{equation}%
under a simple point transformation assumption  
\begin{equation}
q=q\left( x\right) =\int \text{ }\sqrt{m\left( x\right) }dx=\sqrt{Q\left(
x\right) }x\Longleftrightarrow \text{ }\sqrt{m\left( x\right) }=\sqrt{%
Q\left( x\right) }\left( 1+\frac{Q^{\prime }\left( x\right) }{2Q\left(
x\right) }x\right) .
\end{equation}%
This would, in turn, with the PDM of (13) imply that 
\begin{equation}
Q\left( x\right) =\frac{1}{\lambda x^{2}}\ln \left( \sqrt{\lambda }x+\sqrt{%
\lambda x^{2}+1}\right) ^{2}; \lambda>0,
\end{equation}%
and hence the corresponding PDM standard potential which adapts itself to
accommodate the PDM settings of (13) is given by%
\begin{equation}
V\left( x\right) =\frac{1}{2}\left[ \frac{\omega ^{2}}{\lambda }\ln \left( 
\sqrt{\lambda }x+\sqrt{\lambda x^{2}+1}\right) ^{2}\right] ,
\end{equation}%
\begin{figure}[h!]  
\centering
\includegraphics[width=0.3\textwidth]{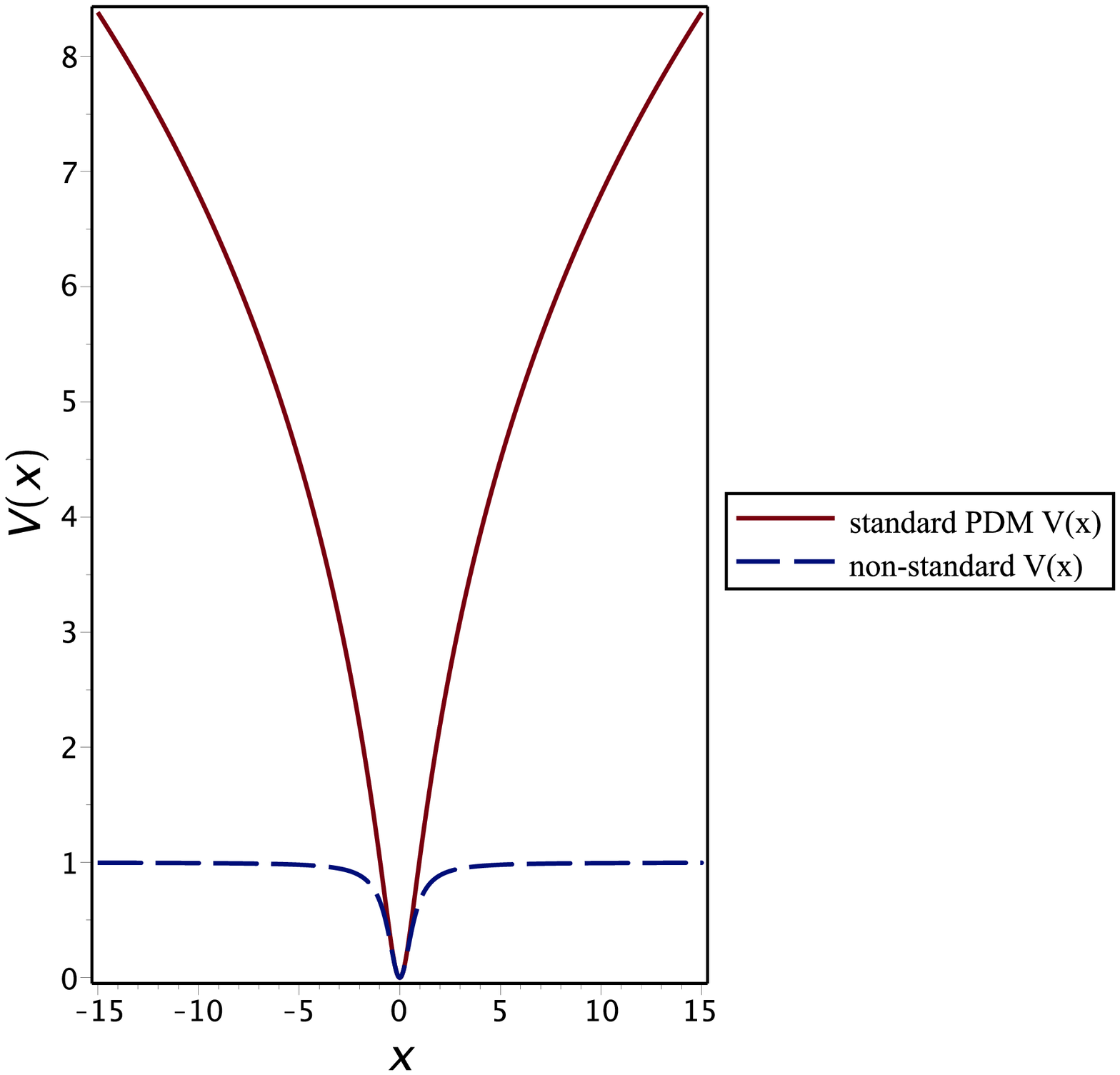}
\includegraphics[width=0.3\textwidth]{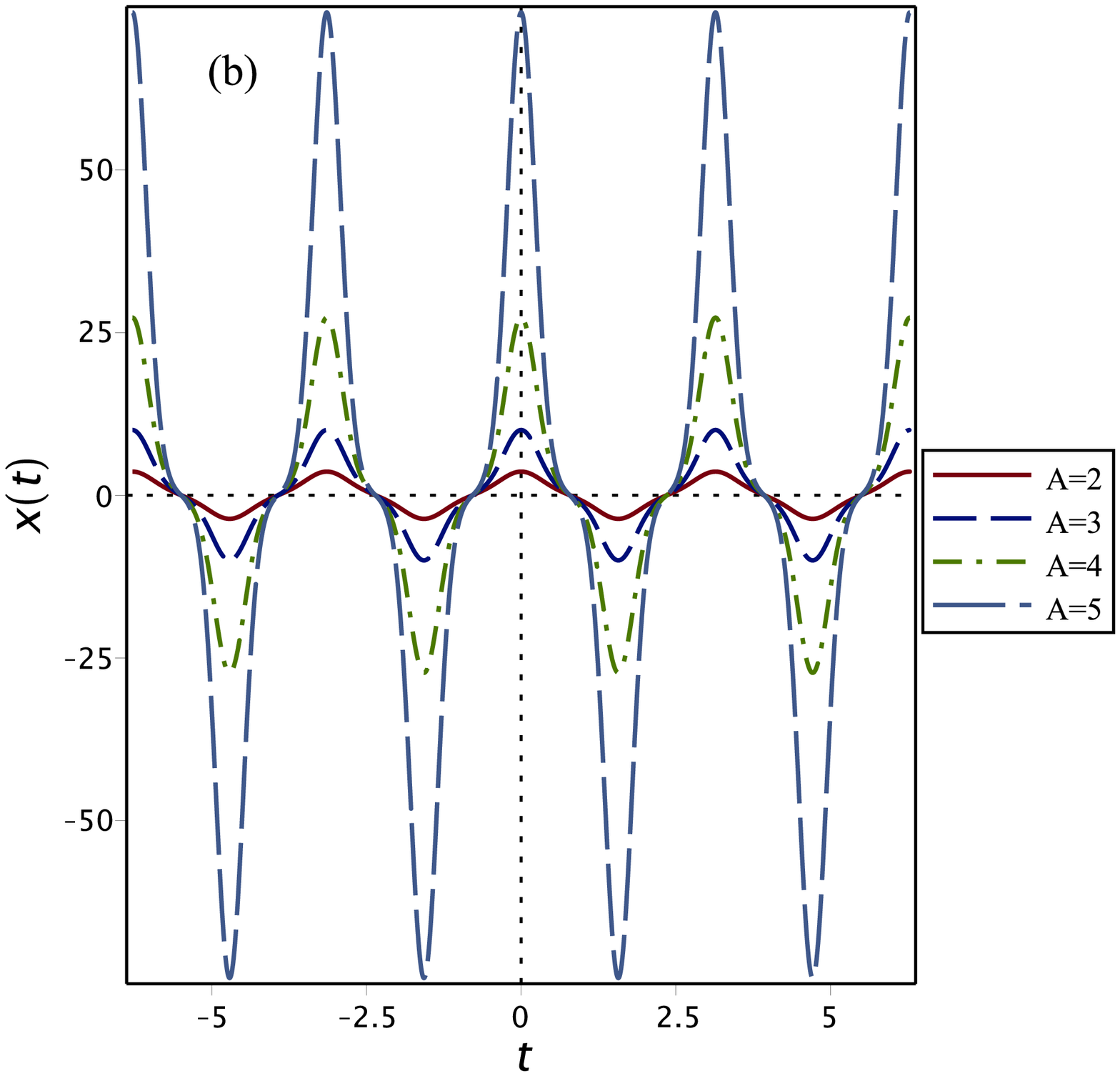} 
\includegraphics[width=0.3\textwidth]{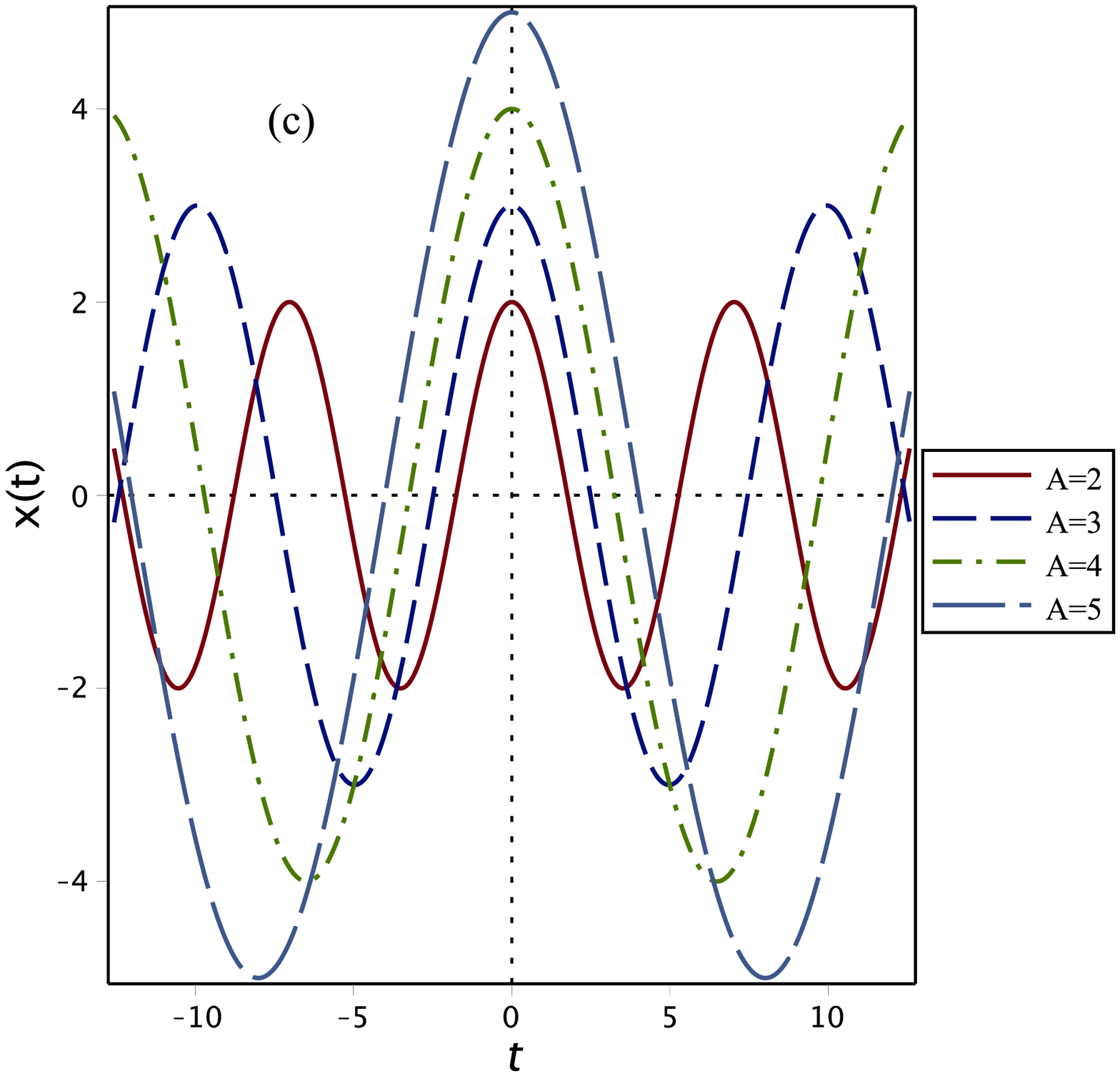}
\caption{\small 
{Shows (a) the standard potential energy (19) and, (b) the non-standard potential of (8), (b) Isochronous oscillator of the standard dynamical equation (20), and (c) the non-isochronous oscillator of the non-standard dynamical equation (6).}}
\label{fig1}
\end{figure}
Then the corresponding standard PDM dynamical equation reads%
\begin{equation}
\ddot{x}-\frac{\lambda x}{1+\lambda x^{2}}\dot{x}^{2}+\frac{\omega ^{2}}{%
\lambda }\sqrt{\lambda x^{2}+1}\ln \left( \sqrt{\lambda }x+\sqrt{\lambda
x^{2}+1}\right) =0
\end{equation}%
Which admits exact solution%
\begin{equation}
q=\sqrt{Q\left( x\right) }x\Longleftrightarrow x=\frac{1}{2\lambda }\left(
e^{\lambda A\cos \left( \omega t+\varphi \right) }-e^{-\lambda A\cos \left(
\omega t+\varphi \right) }\right) .
\end{equation}%
In Figure 1(a) we show the standard PDM potential (19) and the non-standard ML
potential of (8) corresponding to the PDM particle of (13). We observe that
the non-standard ML-potential may allow simple harmonic motion within a very
narrow region in space (restricted via the amplitude dependent frequencies
in (7)), whereas the standard PDM potential (19) offers a simple harmonic
motion within the full space. The ML-potential flattens out into $V(x)=\omega^{2}/2\lambda$ as $x\longrightarrow\pm\infty$. Moreover, we show, in Figure 1(b), the
isochronuos behavior of the standard PDM oscillator solution (21)
(corresponding to the standard dynamical equation in (20)), and the
non-isochronuos, in Figure 1(c), behavior of ML-oscillator (7) (corresponding to the
non-standard dynamical equation in (6)). The phase trajectories are also
reported in Figures 2(a) and 2(b) documenting a narrow region in space for the non-standard non-isochronuos ML oscillator (using the same parametric setting).
\begin{figure}[h!]  
\centering
\includegraphics[width=0.4\textwidth]{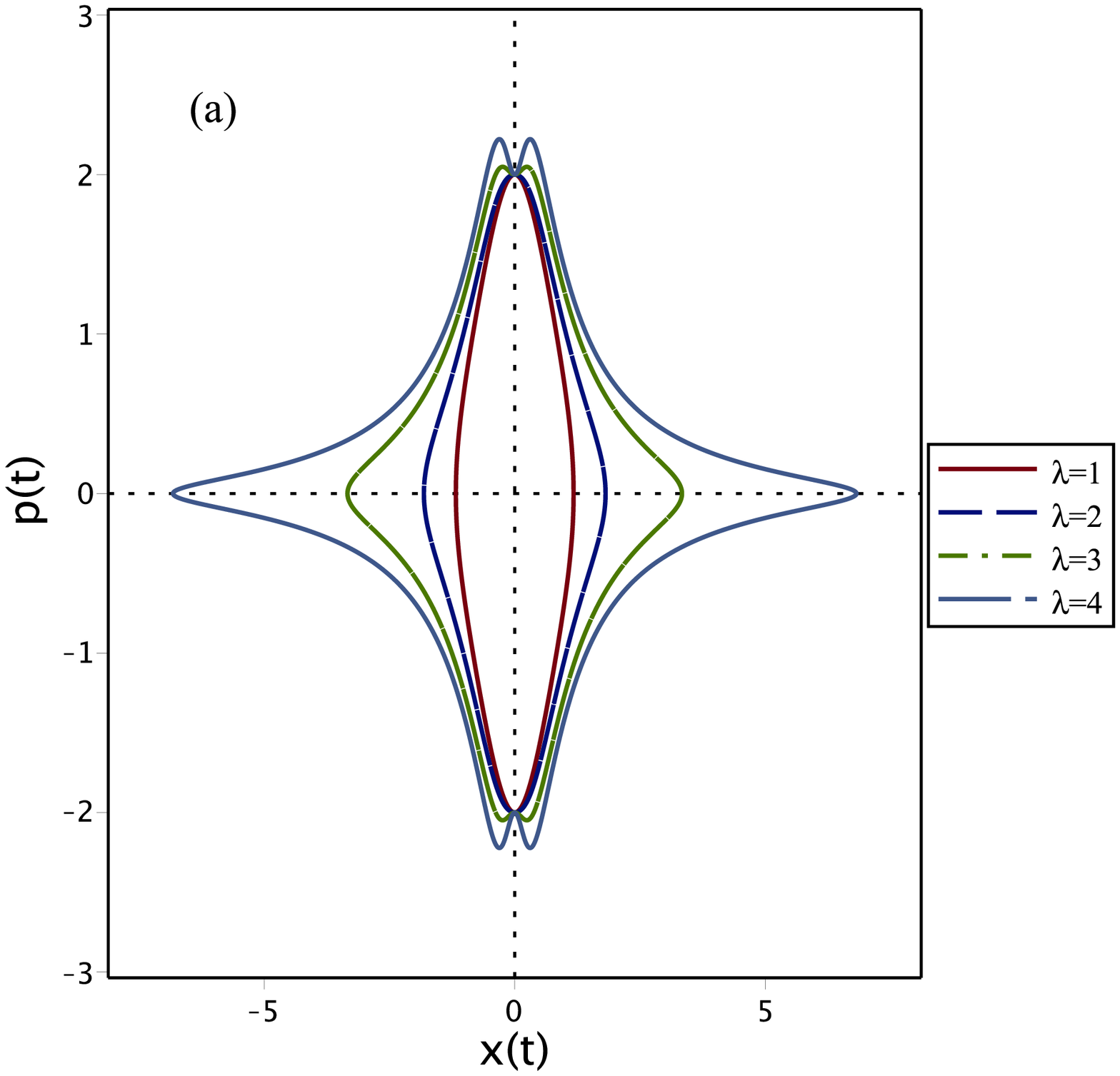} 
\includegraphics[width=0.4\textwidth]{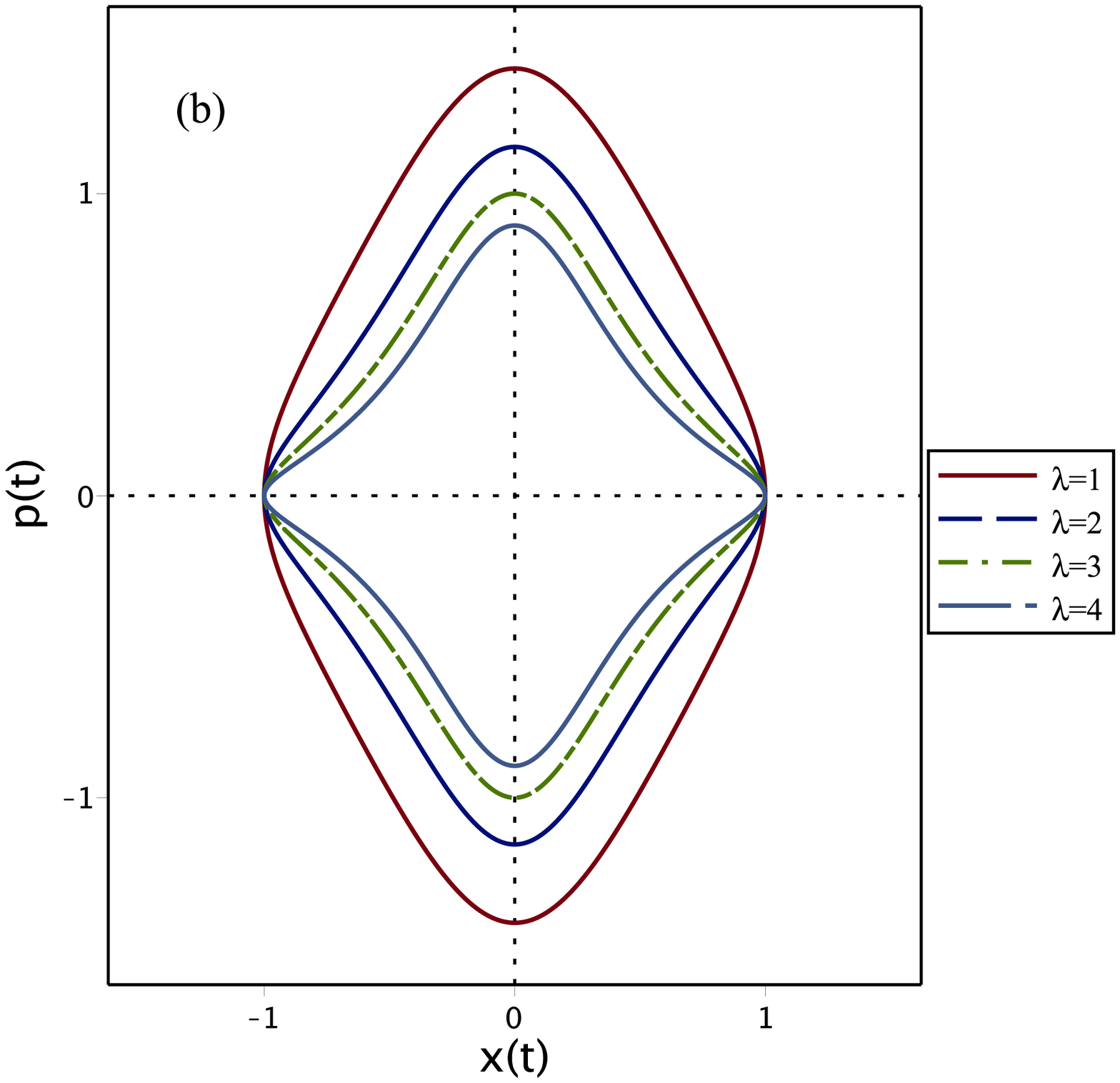}
\caption{\small 
{Shows the Phase trajectories for (a) the Isochronous oscillator of the standard dynamical equation (20), and (b) the non-isochronous oscillator (7) of the non-standard dynamical equation (6).}}
\label{fig2}
\end{figure}

Unavoidably, at this point, one should mention that the linearizability of
the dynamical system of (11) into a similar form of (16) may be achieved
through two nonlocal point transformations (to the best of our knowledge).
The first of which (c.f., e.g., \cite{3,4} and related references cited therein) suggests that%
\begin{equation}
du=\sqrt{m\left( x\right) }dx\,\text{\ \ };\text{ \ }d\tau =m\left( x\right)
\,dt\Longleftrightarrow \frac{du}{d\tau }=\frac{1}{m\left( x\right) }\left(
1+\frac{m^{\prime }\left( x\right) }{2m\left( x\right) }x\right) \sqrt{%
m\left( x\right) }\dot{x}=\sqrt{m\left( x\right) }\dot{x},
\end{equation}%
which is valid if and only if $m\left( x\right) $ satisfies condition (12)
(hence, resulting the PDM in (13)) and consequently imply the dynamical
system in (6). The second nonlocal point transformation, on the other hand,
is a more general one and applies for all $m\left( x\right) $ (e.g., \cite%
{5} for more details on this issue) and is given by%
\begin{equation}
du=\sqrt{g\left( x\right) }dx\,,\text{ }d\tau =f\left( x\right) \,dt\,,\text{
}u=\int \sqrt{g\left( x\right) }dx=\sqrt{m\left( x\right) }x.
\end{equation}%
Which, in a straightforward manner, yields the dynamical equation%
\begin{equation}
\ddot{x}+\frac{m^{\prime }\left( x\right) }{2m\left( x\right) }\dot{x}%
^{2}+f\left( x\right) \omega ^{2}x=0;\;\sqrt{g\left( x\right) }=f\left(
x\right) \sqrt{m\left( x\right) },\;\,f\left( x\right) =\left( 1+\frac{%
m^{\prime }\left( x\right) }{2m\left( x\right) }x\right) .
\end{equation}%
For which,  $f\left( x\right) =m\left( x\right) $ necessarily results in $%
m\left( x\right) =1/\left( 1+\lambda x^{2}\right) $. Hence, equation (24)
collapses into (6) and admits the non-isochronous simple harmonic oscillator
solution of (7). Notably, in the two nonlocal point transformations,  (22)
and (23), the time element $d\tau $ is rendered position dependent (i.e., $%
d\tau =m\left( x\right) \,dt$ in (22) and $d\tau =f\left( x\right) \,dt$ in
(23)). This would naturally and manifestly affect the periods of
oscillations and consequently the frequencies of oscillations become
amplitude dependent ones, i.e., non-isochronuos. Therefore, if the
oscillators isochronicity is the sought after objective then the time
element should not be a position-dependent deformed one. Isochronous
PDM-oscillators form the focal point of our forthcoming $n$-dimensional
study. 

\section{$n$-dimensional gradient of PDM-potential and invariance}

Apriori, it is known that under constant mass setting, the force is the time
derivative of the canonical momentum and is given by negative the gradient
of the potential force field., i. e., 
\begin{equation}
\mathbf{F=}\frac{d\mathbf{p}}{dt}=-\mathbf{\nabla }V(\mathbf{r})\,;\mathbf{%
\nabla }=\sum\limits_{j=1}^{3}\partial _{x_{_{j}}}\,\hat{x}_{_{j}}\,,\,%
\mathbf{r}=\sum\limits_{j=1}^{3}x_{_{j}}\,\hat{x}_{_{j}}\,,\,\mathbf{F=}%
\sum\limits_{j=1}^{3}F_{_{i}}\hat{x}_{_{j}}\,,\,r=\sqrt{\sum%
\limits_{j=1}^{3}x_{_{j}}^{2}}.
\end{equation}%
Under PDM settings, however, negative the gradient of the potential force
field is no longer given by the time derivative of the canonical momentum.
In the one-dimensional case, for example, Mustafa \cite{5} has asserted that
if the PDM potential term in Lagrangian (10) is replaced by its PDM form $V\left( x\right)$ then  the relation
between the force and the potential force field is rather given by (3) as%
\begin{equation}
F=m\left( x\right) \,\ddot{x}+\frac{m^{\prime }\left( x\right) }{2}\dot{x}%
^{2}=\sqrt{m\left( x\right) }\frac{d}{dt}\left( \sqrt{m\left( x\right) }\dot{%
x}\right) =-\frac{d}{dx}V\left( x\right) \,
\end{equation}%
where $V\left( x\right) =V\left( q\left( x\right) \right) $ is the PDM-deformed potential force field, and $q\left( x\right) $ is given by
(17). It is obvious that equation (17) gives the relation between the
deformation $\sqrt{Q\left( x\right) }$ in the coordinate $x$ and the
deformation $\sqrt{m\left( x\right) }$ in the velocity $\dot{x}$. Moreover,
equation (26) is a documentation that, in the one-dimensional case, negative
the gradient of the potential force field is not equal to the time
derivative of the canonical momentum (i.e., $dp\mathbf{/}dt\neq -V^{\prime
}(x),\,$where $p=m\left( x\right) \dot{x}$). So is the $n$-dimensional case.
Consequently, the underlying $n$-dimensional dynamics of the PDM systems
have to be clarified. Namely, one has to answer the question as to "what
would negative the gradient of the $n$-dimensional PDM-potential force field
yield to? That would be the sought after net PDM force vector.

\subsection{Negative the gradient of the PDM potential force field}

Consider the $n$-dimensional PDM Lagrangian%
\begin{equation}
L\left( \mathbf{r},\mathbf{\dot{r}};t\right) =\frac{1}{2}m_{\circ }m\left( 
\mathbf{r}\right) \,\mathbf{\dot{r}}^{2}-V\left( \mathbf{r}\right) =\frac{1}{%
2}m_{\circ }m\left( \mathbf{r}\right) \sum\limits_{j=1}^{n}\dot{x}%
_{_{j}}^{2}-V\left( \mathbf{r}\right) \,,\text{ }
\end{equation}%
where $V\left( \mathbf{r}\right) =V\left( \mathbf{q}\left( \mathbf{r}\right)
\right) $ is an $n$-dimensional PDM deformed potential force field and  
\begin{equation}
\mathbf{q}\left( \mathbf{r}\right) =\sqrt{Q\left( \mathbf{r}\right) }\mathbf{%
r=}\sum\limits_{j=1}^{n}q_{_{j}}\left( \mathbf{r}\right) \hat{x}%
_{_{j}}\Longleftrightarrow q_{_{j}}\left( \mathbf{r}\right) =\sqrt{Q\left( 
\mathbf{r}\right) }x_{_{j}},
\end{equation}%
serves as the $n$-dimensional generalization of (17). As long as the
relation between $m\left( \mathbf{r}\right) $ and $Q\left( \mathbf{r}\right) 
$ is to be determined in the process, this assumption remains valid and
sufficient. We may now use the $n$-dimensional Euler-Lagrange equations 
\begin{equation}
\frac{d}{dt}\left( \frac{\partial L}{\partial \dot{x}_{i}}\right) -\frac{%
\partial L}{\partial x_{i}}=0;\text{ }\,\,\,i=1,2,\cdots ,n\in 
\mathbb{N}
,
\end{equation}%
to obtain (with $m_{\circ }=1$) $n$ Euler-Lagrange equations%
\begin{equation}
m\left( \mathbf{r}\right) \,\ddot{x}_{_{i}}+\dot{m}\left( \mathbf{r}\right) 
\dot{x}_{i}-\frac{1}{2}\partial _{x_{i}}m\left( \mathbf{r}\right)
\sum\limits_{j=1}^{n}\dot{x}_{_{j}}^{2}=-\,\partial _{x_{i}}V\left( \mathbf{r%
}\right) .
\end{equation}%
Next, we multiply each term by the related unit vector $\mathbf{\hat{x}}_{_{i}}$ and
sum over $i=1,2,\cdots ,n$ to get the corresponding Newtonian
dynamical equation in its vector formation
\begin{equation}
m\left( \mathbf{r}\right) \sum\limits_{i=1}^{n}\ddot{x}_{_{i}}\,\mathbf{\hat{x}}_{_{i}}+\dot{m}\left( \mathbf{r}\right) \sum\limits_{i=1}^{n}\dot{x}_{_{i}}\,%
\mathbf{\hat{x}}_{_{i}}-\frac{1}{2}\sum\limits_{i=1}^{n}\partial _{x_{i}}m\left( 
\mathbf{r}\right)\mathbf{\hat{x}}_{_{i}}\left[ \sum\limits_{j=1}^{n}\dot{x}%
_{_{j}}^{2}\right] =-\sum\limits_{i=1}^{n}\mathbf{\hat{x}}_{_{i}}\partial
_{x_{i}}V\left( \mathbf{r}\right)=-\mathbf{\nabla }V\left( \mathbf{r}%
\right) .
\end{equation}%
To avoid mathematical complexities, we may assume that $m\left( \mathbf{r}%
\right) =m\left( r\right) $ and $Q\left( \mathbf{r}\right) =Q\left( r\right) 
$ where $r$ is readily defined in (25). This would allow us to represent
(31) as%
\begin{equation}
m\left( r\right) \,\mathbf{\ddot{r}}+\dot{m}\left( r\right) \,\mathbf{\dot{r}%
}-\frac{1}{2}{\nabla }m\left( r\right)%
{\dot{r}}^{2}=-\mathbf{\nabla }V\left( \mathbf{r}\right) .
\end{equation}%
However, one may express $\dot{m}\left( r\right) $, with $m^{\prime }\left(
r\right) =\partial m\left( r\right) /\partial r$, as%
\begin{equation}
\dot{m}\left( r\right) =\sum\limits_{k=1}^{n}\partial _{x_{k}}m\left(
r\right) \,\dot{x}_{_{k}}=\frac{m^{\prime }\left( r\right) }{r}%
\sum\limits_{k=1}^{n}x_{_{k}}\dot{x}_{_{k}}=\frac{m^{\prime }\left( r\right) 
}{r}\,\left( \mathbf{r}\cdot \mathbf{\dot{r}}\right),
\end{equation}%
and%
\begin{equation}
{\nabla }m\left( r\right)=\sum\limits_{i=1}^{n}\partial _{x_{i}}m\left( r\right) \,\mathbf{\hat{x}}_{_{i}}=%
\frac{m^{\prime }\left( r\right) }{r}\sum\limits_{k=1}^{n}x_{_{k}}\mathbf{\hat{x}}_{_{k}}=\frac{m^{\prime }\left( r\right) }{r}\,\mathbf{r},
\end{equation}%
so that equation (32) reads, with $\mathbf{r\,}{\dot{r}}^{2}=\mathbf{r\,}\left( \mathbf{\dot{r}\cdot 
\dot{r}}\right) =\,\left( \mathbf{r}\cdot \mathbf{\dot{r}}\right) \,\mathbf{%
\dot{r}}$ (i.e., no rotational effects under consideration and $\mathbf{r}%
\parallel \mathbf{\dot{r}}$, therefore), 
\begin{equation}
m\left( r\right) \,\mathbf{\ddot{r}}+\frac{\dot{m}\left( r\right) }{2}\,%
\mathbf{\dot{r}}+\mathbf{\nabla }V\left( \mathbf{r}\right)
=0\Longleftrightarrow \mathbf{F}=\sqrt{m\left( r\right) }\frac{d}{dt}\left( 
\sqrt{m\left( r\right) }\mathbf{\dot{r}}\right) =-\mathbf{\nabla }V\left( 
\mathbf{r}\right) .
\end{equation}%
This result would, in fact, represent the $n$-dimensional PDM Newtonian
dynamical equations in their total vector format. Yet, the procedure
described in (30) to (35) identifies the transition from Euler-Lagrange
description of the vector-components into the total vector structure of Newtonian
dynamical equations. The advantage of such total vector description has been
discussed in [42] and shall be recollected (to make the current study self-contained) in the forthcoming section.
Moreover, this result (35) suggests that in a free force field (i.e., $%
V\left( \mathbf{q}\left( \mathbf{r}\right) \right) =0$), the canonical
momentum $\mathbf{p}=m\left( r\right) \mathbf{\dot{r}}$ is no longer a
conserved quantity but rather the PDM pseudo-momentum vector $\mathbf{\pi} \left(\mathbf{
r}\right) =\sqrt{m\left( r\right) }\mathbf{\dot{r}}$ \cite{5,7} (or in the
Cari\v{n}ena et al's \cite{6} language, the "Noether momentum") is the
conserved quantity. It is now obvious that, under PDM setting, negative the
gradient of the potential force field is no longer the same as the time
derivative of the canonical momentum $\mathbf{p}$. Yet it recovers the
constant mass settings for $m\left( r\right) =1$ to yield the usual textbook
relation $m_{\circ }\,\mathbf{\ddot{r}}=-\mathbf{\nabla }V\left( \mathbf{r}%
\right) $.

\subsection{$n$-dimensional PDM Lagrangians: point transformation and  invariance}

Let us consider a standard $n$-dimensional constant mass Lagrangian%
\begin{equation}
L\left( \mathbf{q},\mathbf{\dot{q}};t\right) =\frac{1}{2}m_{\circ
}\sum\limits_{j=1}^{n}\dot{q}_{_{j}}^{2}-V(\mathbf{q});\text{ \ }\dot{q}%
_{_{j}}=\frac{dq_{_{j}}}{dt};\text{ }\,j=1,2,\cdots ,n\in 
\mathbb{N}
,
\end{equation}%
Then the corresponding $n$ Euler-Lagrange equations (with $m_{\circ }=1$)
are given by%
\begin{equation}
\ddot{q}_{_{i}}+\partial _{q_{_{i}}}V(\mathbf{q})=0\text{ };\text{ }%
i=1,2,\cdots,n\in 
\mathbb{N}
.
\end{equation}%
Under a point transformation in the form of 
\begin{equation}
dq_{_{i}}=\sqrt{m\left( r\right) }dx_{_{i}}\,\Longleftrightarrow \partial
_{x_{_{i}}}q_{_{i}}=\,\frac{\partial q_{_{i}}}{\partial x_{_{i}}}=\sqrt{%
m\left( r\right) }\Longleftrightarrow \dot{q}_{_{i}}=\sqrt{m\left( r\right) }%
\dot{x}_{_{i}}\Longleftrightarrow \mathbf{\dot{q}=}\sqrt{m\left( r\right) }%
\,\,\mathbf{\dot{r}},
\end{equation}%
and the assumption (28) that%
\begin{equation}
\mathbf{q}=\sqrt{Q\left( r\right) }\mathbf{r}\Longleftrightarrow \mathbf{%
\dot{q}=}\sqrt{Q\left( r\right) }\left( 1+\frac{Q^{\prime }\left( r\right) }{%
2Q\left( r\right) }r\right) \,\,\mathbf{\dot{r}},
\end{equation}%
one obtains, through the comparison between (38) and (39), that%
\begin{equation}
\sqrt{m\left( r\right) }=\sqrt{Q\left( r\right) }\left( 1+\frac{Q^{\prime
}\left( r\right) }{2Q\left( r\right) }r\right) ;\text{ }r=\sqrt{%
\sum\limits_{j=1}^{n}x_{_{j}}^{2}}.
\end{equation}%
Which is, in fact, analogous to that in (17) . The  correlation between $%
m\left( r\right) $ and $Q\left( r\right) $ is clear, therefore. We may now
proceed with (37) and use%
\begin{equation}
\dot{q}_{_{i}}=\sqrt{m\left( r\right) }\dot{x}_{_{i}}\Longleftrightarrow 
\ddot{q}_{_{i}}=\sqrt{m\left( r\right) }\left[ \ddot{x}_{_{i}}+\frac{\dot{m}%
\left( r\right) }{2m\left( r\right) }\dot{x}_{i}\right] 
\end{equation}%
of (38), along with $\partial _{q_{_{i}}}=\left( \partial x_{_{i}}/\partial
q_{_{i}}\right) \partial _{x_{_{i}}}=m\left( r\right) ^{-1/2}\partial
_{x_{_{i}}}$, to obtain%
\begin{equation}
m\left( r\right) \,\ddot{x}_{_{i}}+\frac{1}{2}\dot{m}\left( r\right) \dot{x}%
_{i}+\,\partial _{x_{i}}V\left( \mathbf{q}\right) =0
\end{equation}%
Which, when compared with (30), suggests that the Euler-Lagrange invariance
between (30) and (37) is still far beyond reach at this stage. Hence,
alternative types of invariance have to be sought at this point. 

Let us multiply equation (42) by the corresponding unit vectors $\mathbf{\hat{x}_{_{i}}}$
and sum over $i=1,2,\cdots,n$ to get the corresponding Newtonian
dynamical equation in total vector presentation%
\begin{equation}
\sum\limits_{i=1}^{n}m\left( r\right) \,\ddot{x}_{_{i}}\,\mathbf{\hat{x}_{_{i}}}+\frac{1%
}{2}\dot{m}\left( r\right) \sum\limits_{i=1}^{n}\dot{x}_{i}\,\mathbf{\hat{x}%
_{_i}}+\sum\limits_{i=1}^{n}\partial _{q_{_{i}}}V(\mathbf{q})\,\mathbf{\hat{x}%
_{_i}}=0\Longleftrightarrow m\left( r\right) \,\mathbf{\ddot{r}}+\frac{\dot{m}%
\left( r\right) }{2}\,\mathbf{\dot{r}}+\mathbf{\nabla }V\left( \mathbf{q}%
\right) =0.
\end{equation}%
Obviously, this result is in exact accord with (35), with $V(\mathbf{q}) =V( \mathbf{r})$, and documents Newtonian invariance between the
two dynamical systems of (27) and (36) (i.e., (35) and (42), respectively). That is,%
\begin{equation}
\sum\limits_{i=1}^{n}\left[ \frac{d}{dt}\left( \frac{\partial L}{\partial 
\dot{x}_{i}}\right) -\frac{\partial L}{\partial x_{i}}\right] \mathbf{\hat{x}%
_{_i}}=0=\sum\limits_{i=1}^{n}\left[ \frac{d}{dt}\left( \frac{\partial L}{%
\partial \dot{q}_{i}}\right) -\frac{\partial L}{\partial q_{i}}\right] \mathbf{\hat{x%
}_{_i}},
\end{equation}%
Which is called \textit{Newtonian invariance amendment} (for it is just the
total vector representation of the Euler-Lagrange dynamical equations)
introduced by Mustafa [42], who have used the nonlocal point transformation
(24) for some non-standard Lagrangians. Nevertheless, for the sake of
completeness of the current methodical proposal, we report that one may very
well seek yet another type of invariance through the time derivative of the
corresponding Hamiltonian systems. That is, one may, in a straightforward
manner show that%
\begin{equation}
\frac{d}{dt}\left( \frac{1}{2}\mathbf{P}^{2}+V(%
\mathbf{q})\right) =0=\frac{d}{dt}\left( \frac{\mathbf{p}^{2}}{2m\left(
r\right) }+V\left( \mathbf{r}\right) \right) \Longleftrightarrow \dot{H}%
\left( \mathbf{q},\mathbf{P};t\right) =0=\dot{H}\left( \mathbf{r},%
\mathbf{\dot{p}};t\right) ,
\end{equation}
where $\mathbf{P}=\mathbf{\dot{q}}$ and $\mathbf{p}=m\left( r\right) \mathbf{\dot{r}}$ are the corresponding canonical momenta for the two systems. Such types of invariance,
however, give us the authority to use the exact solutions of one system and
map it, with ease, into the other. This would consequently enrich the class
of exactly solvable dynamical systems within the standard
Lagrangian/Hamiltonian settings.

\section{Isochronous $n$-dimensional PDM oscillators:
linearizability and invariance}

Having had settled down the technical mathematical issues in the preceding
sections, we may now proceed to discuss the $n$-dimensional PDM harmonic
oscillators linearizability, invariance and isochronicity. We begin with the 
$n$-dimensional PDM oscillator Lagrangian%
\begin{equation}
L\left( \mathbf{r},\mathbf{\dot{r}};t\right) =\frac{1}{2}m\left( r\right) \,%
\mathbf{\dot{r}}^{2}-V\left( \mathbf{r}\right) =\frac{1}{2}m\left( r\right)
\sum\limits_{j=1}^{n}\dot{x}_{_{j}}^{2}-\frac{1}{2}\omega ^{2}Q\left(
r\right) \sum\limits_{j=1}^{n}x_{_{j}}^{2},
\end{equation}%
where the oscillator potential is now assumed to be PDM-deformed in such a
way that $\mathbf{r}\longrightarrow\sqrt{Q\left( r\right) }\mathbf{r}$ as a
consequence of the PDM settings discussed above. The substitution of the
PDM oscillator Lagrangian (46) in the $n$ Euler-Lagrange equations of motion
(35) would result the Euler-Lagrange dynamical equations, in the vector components form, 
\begin{equation}
\ddot{x}_{_{i}}+\frac{\dot{m}\left( r\right) }{m\left( r\right) }\dot{x}_{i}-%
\frac{1}{2}\partial _{x_{i}}m\left( \mathbf{r}\right) \sum\limits_{j=1}^{n}%
\dot{x}_{_{j}}^{2}+\sqrt{\frac{Q\left( r\right) }{m\left( r\right) }}\omega
^{2}x_{i}=0,
\end{equation}%
where we have used the relation (38) in the process. On the other hand, the $%
n$-dimensional constant mass oscillator Lagrangian%
\begin{equation}
L\left( \mathbf{q},\mathbf{\dot{q}};t\right) =\frac{1}{2}\,\mathbf{\dot{q}}%
^{2}-\frac{1}{2}\omega ^{2}\mathbf{q}^{2}=\frac{1}{2}\sum\limits_{j=1}^{n}%
\dot{q}_{_{j}}^{2}-\frac{1}{2}\,\omega ^{2}\sum\limits_{j=1}^{n}q_{_{j}}^{2},
\end{equation}%
yields the $n$ Euler-Lagrange linear differential equations%
\begin{equation}
\ddot{q}_{_{i}}+\omega ^{2}q_{_{i}}=0\text{,}
\end{equation}%
that admit exact sinusoidal oscillatory solutions%
\begin{equation}
q_{_{i}}=A_{_{i}}\cos \left( \omega t+\varphi \right) .
\end{equation}%
Using our point transformation of (38)-(41) in (49) one obtains%
\begin{equation}
\sqrt{m\left( r\right) }\left[ \ddot{x}_{_{i}}+\frac{\dot{m}\left( r\right) 
}{2m\left( r\right) }\dot{x}_{i}\right] +\sqrt{Q\left( r\right) }\omega
^{2}x_{i}=0\Longleftrightarrow \ddot{x}_{_{i}}+\frac{\dot{m}\left( r\right) 
}{2m\left( r\right) }\dot{x}_{i}+\sqrt{\frac{Q\left( r\right) }{m\left(
r\right) }}\omega ^{2}x_{i}=0.
\end{equation}%
This result, again, suggests that, under the current point transformation, the linearizability of (47) into (49) is only possible for the
one-dimensional case. Whereas, for the $n\geq2$ dimensions we observe that the invariance could not be established and the linearization is not feasible. Nevertheless, the two systems are readily invariant, either through the vector totality of Newtonian invariance (44) or $\dot{H}$-invariance (45). This would, in effect, authorize the use of the exact solutions (50) of (49) to find the solutions of (47) (equivalently, solution of (51)). This is illustrated in the sample of examples below.

\section{Isochronous $n$-dimensional nonlinear PDM oscillators: illustrative
examples}

\subsection{One-dimensional isochronous nonlinear PDM oscillators}

For the one-dimensional case one should be aware that the dynamical equations in (47) and (51), associated with the one-dimensional
PDM-oscillators Lagrangians (47)%
\begin{equation*}
L=\frac{1}{2}m\left( x\right) \dot{x}^{2}-\frac{1}{2}Q\left( x\right) \omega
^{2}x^{2},
\end{equation*}
\begin{figure}[h!]  
\centering
\includegraphics[width=0.4\textwidth]{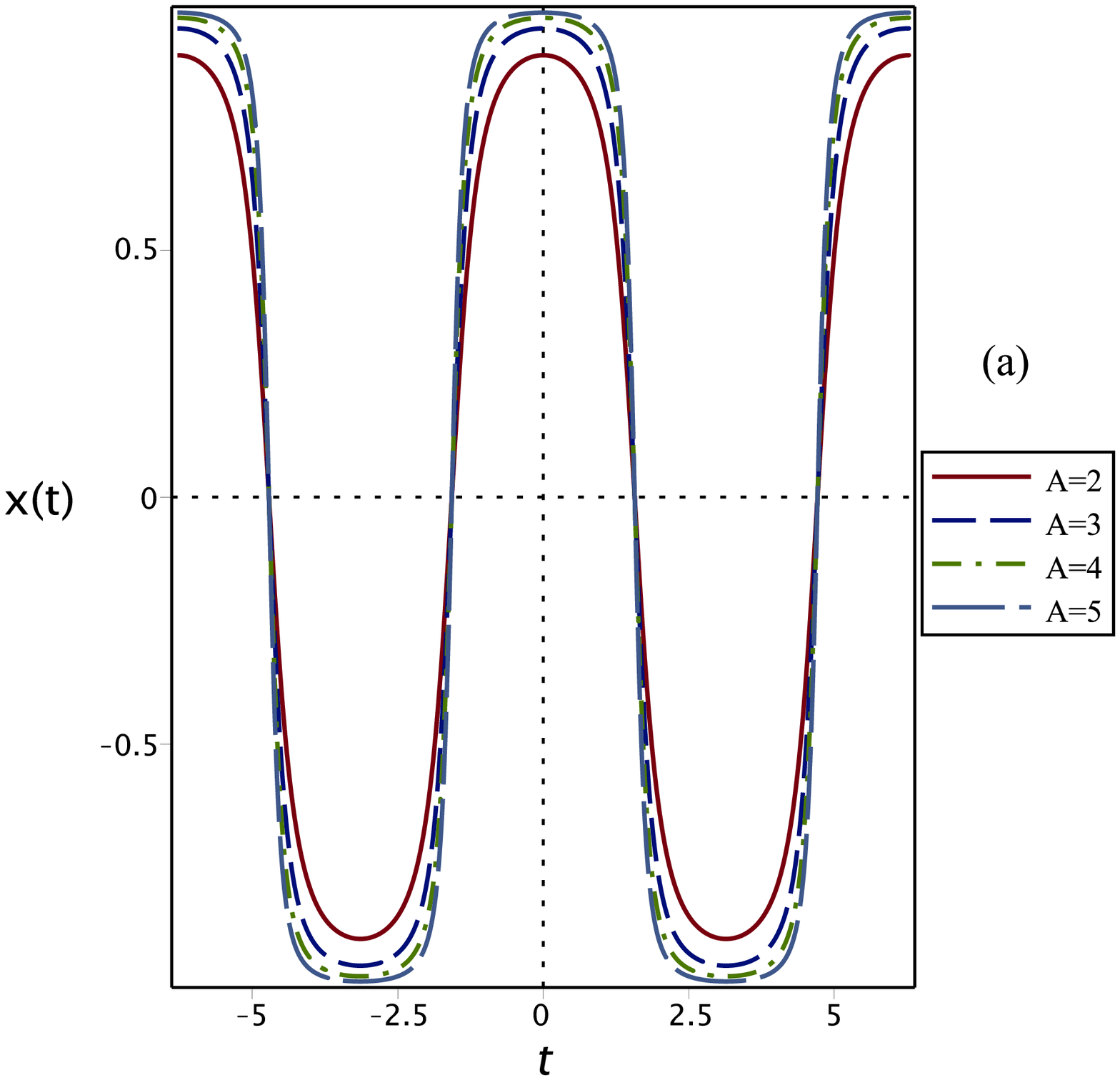} 
\includegraphics[width=0.4\textwidth]{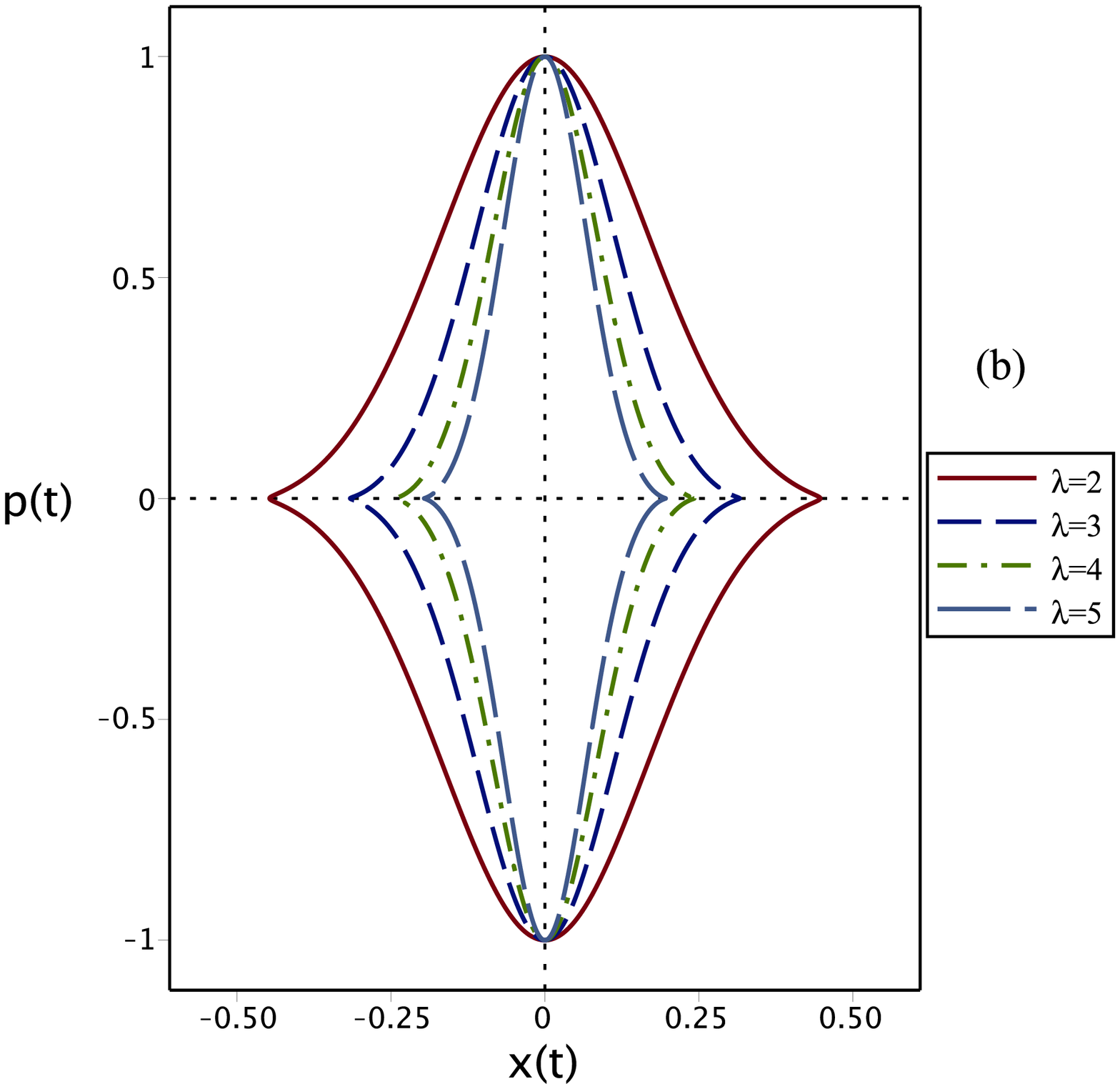}
\caption{\small 
{Shows (a) the Isochronous oscillator of (56), and (b) the Phase trajectories of the Isochronous dynamical system in (55).}}
\label{fig3}
\end{figure}
are identical and the Euler-Lagrange invariance is very well established. Moreover, the linearizability of (47) into (49), in one-dimension, is possible and straightforward.
\subsubsection{A coordinate deformation without singularity: $%
Q\left( x\right) =1/\left( 1+ \lambda ^{2}x^{2}\right) $}

A position-dependent coordinate deformation in the form of%
\begin{equation}
\sqrt{Q\left( x\right) }=\sqrt{\frac{1}{1+\lambda ^{2}x^{2}}},
\end{equation}%
would imply, by (40), a PDM function%
\begin{equation}
m\left( x\right) =\left( \frac{1}{1+\lambda ^{2}x^{2}}\right) ^{3}.
\end{equation}%
Then the dynamical equation (47), or (51), for the one-dimensional PDM-oscillator Lagrangian%
\begin{equation}
L=\frac{1}{2}\left[ \frac{\dot{x}^{2}}{\left( 1+\lambda ^{2}x^{2}\right) ^{3}%
}-\frac{\omega ^{2}x^{2}}{1+\lambda ^{2}x^{2}}\right] ,
\end{equation}%
yields%
\begin{equation}
\ddot{x}-\frac{3\lambda ^{2}x}{1+\lambda ^{2}x^{2}}\dot{x}^{2}+\left(
1+\lambda ^{2}x^{2}\right) \,\omega ^{2}x=0,
\end{equation}%
and admits, using (50) and (39), exact solution in the form of%
\begin{equation}
q=A\cos \left( \omega t+\varphi \right) =\sqrt{Q\left( x\right) }%
x\Longleftrightarrow x=\frac{A\cos \left( \omega t+\varphi \right) }{\sqrt{%
1-\lambda ^{2}A^{2}\cos ^{2}\left( \omega t+\varphi \right) }} ;  0<A<\frac{1}{\lambda}.
\end{equation}%
Which exactly satisfies the dynamical systems in (55) and hence represent their exact isochronous nonlinear PDM-oscillators solutions. In figure 3(a) we show the isochronuous oscillator (56) and in Figure 3(b) we plot the phase trajectory/portrait for the isochronous dynamical PDM system (55).

\subsubsection{A coordinate deformation with a singularity: $Q\left(
x\right) =1/\left( 1-\lambda x\right) $}

A coordinate deformation in the form of%
\begin{equation}
\sqrt{Q\left( x\right) }=\sqrt{\frac{1}{1-\lambda x}},
\end{equation}%
\begin{figure}[h!]  
\centering
\includegraphics[width=0.4\textwidth]{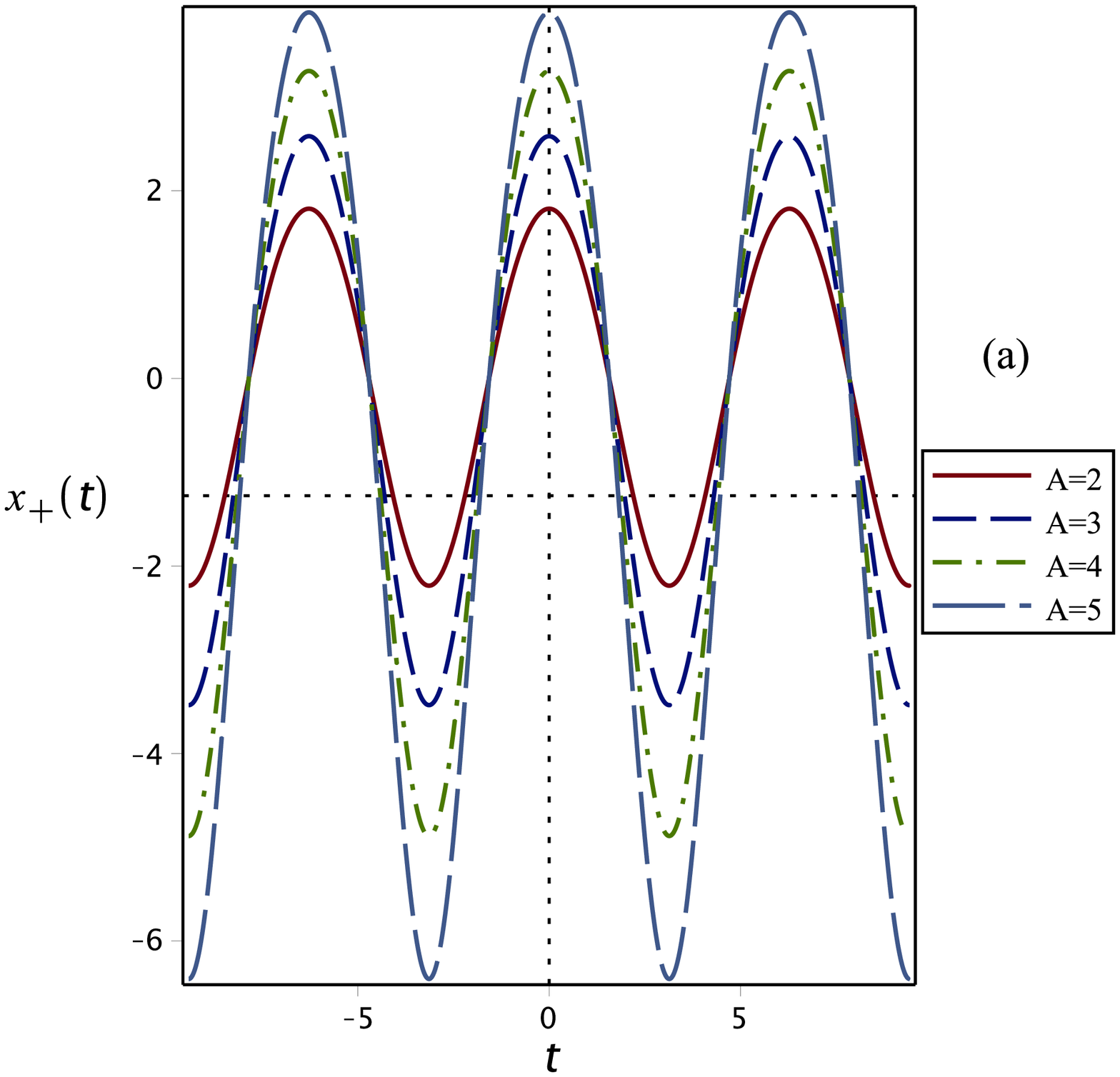} 
\includegraphics[width=0.4\textwidth]{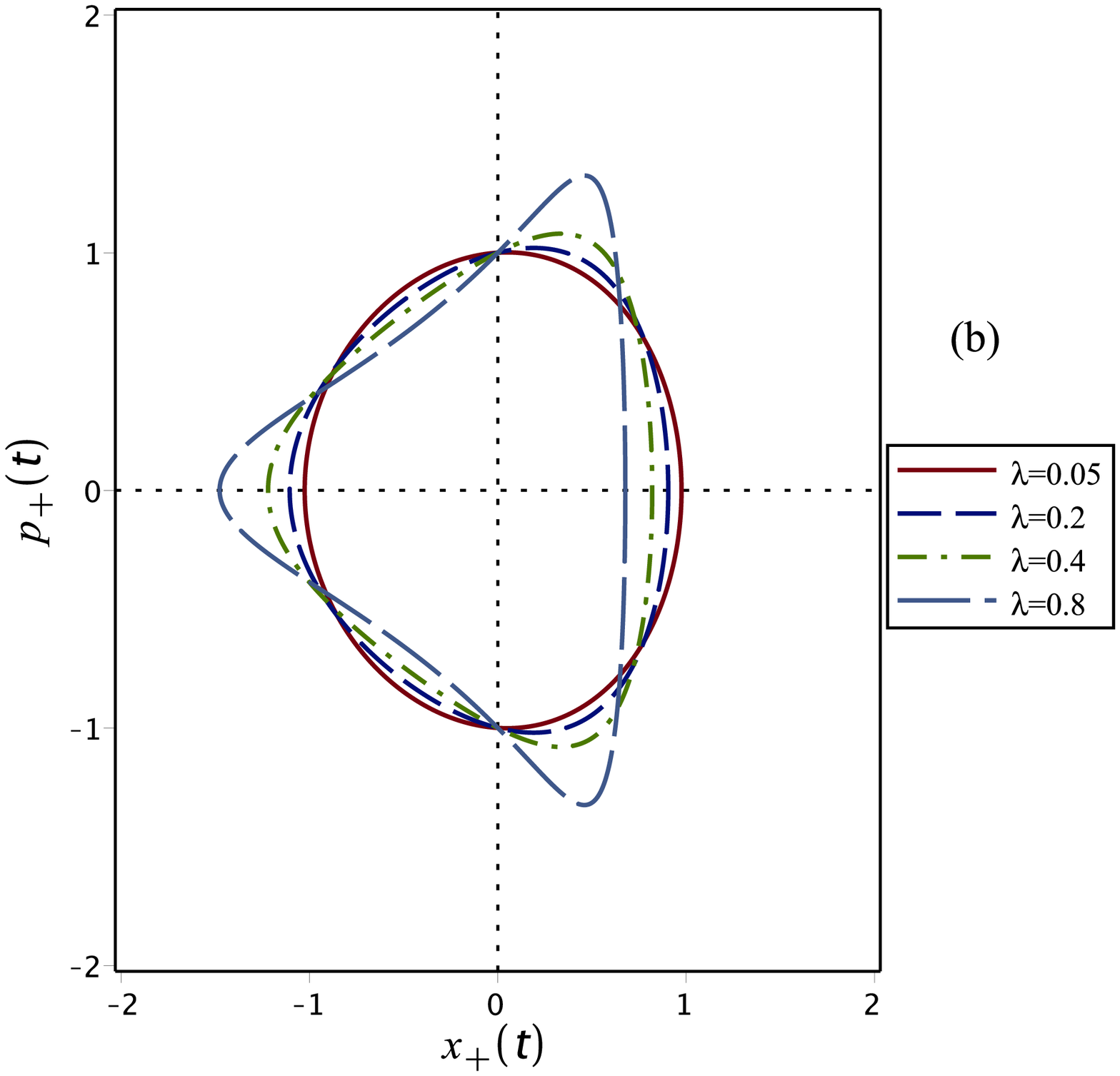}
\includegraphics[width=0.4\textwidth]{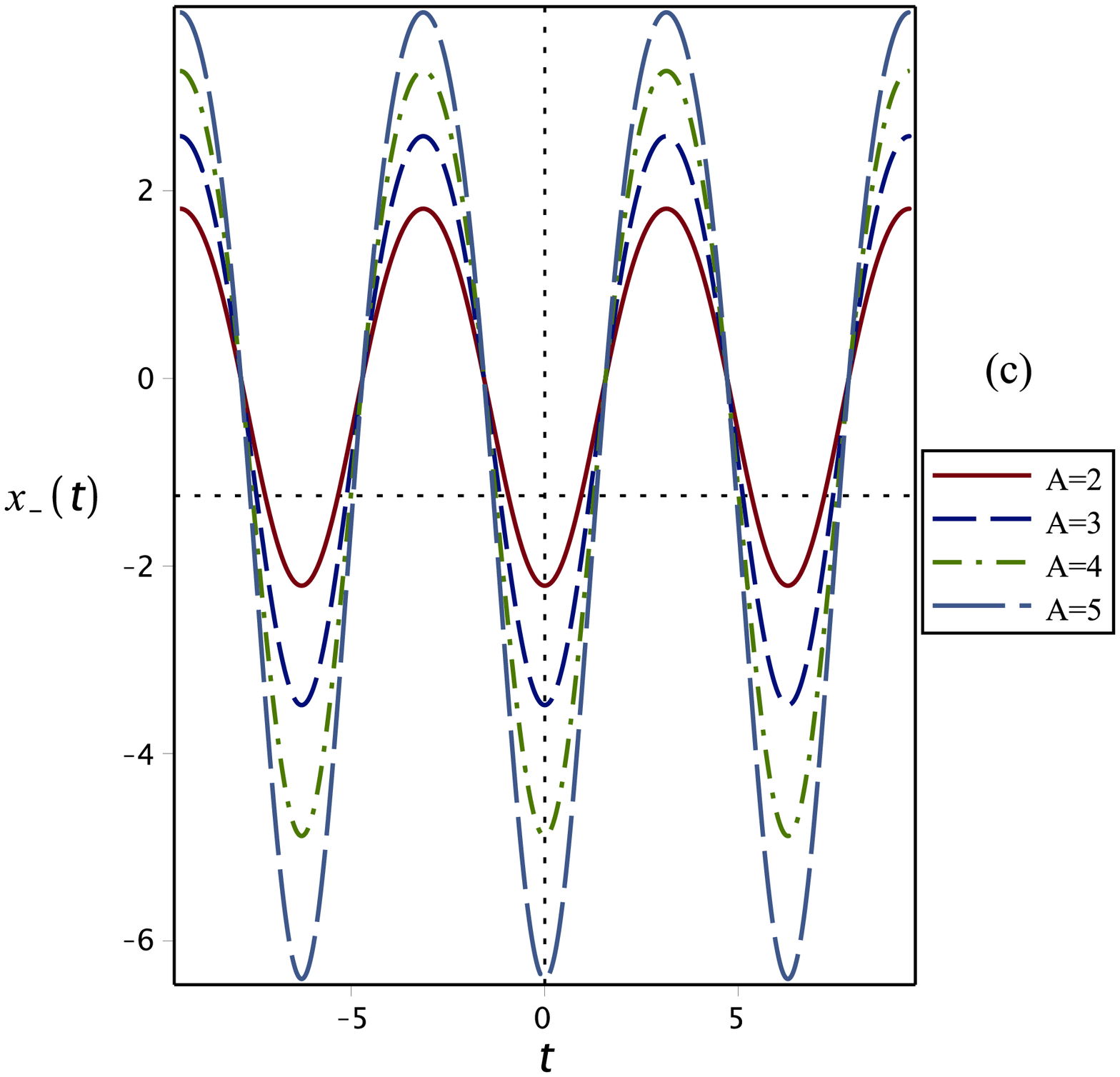}
\includegraphics[width=0.4\textwidth]{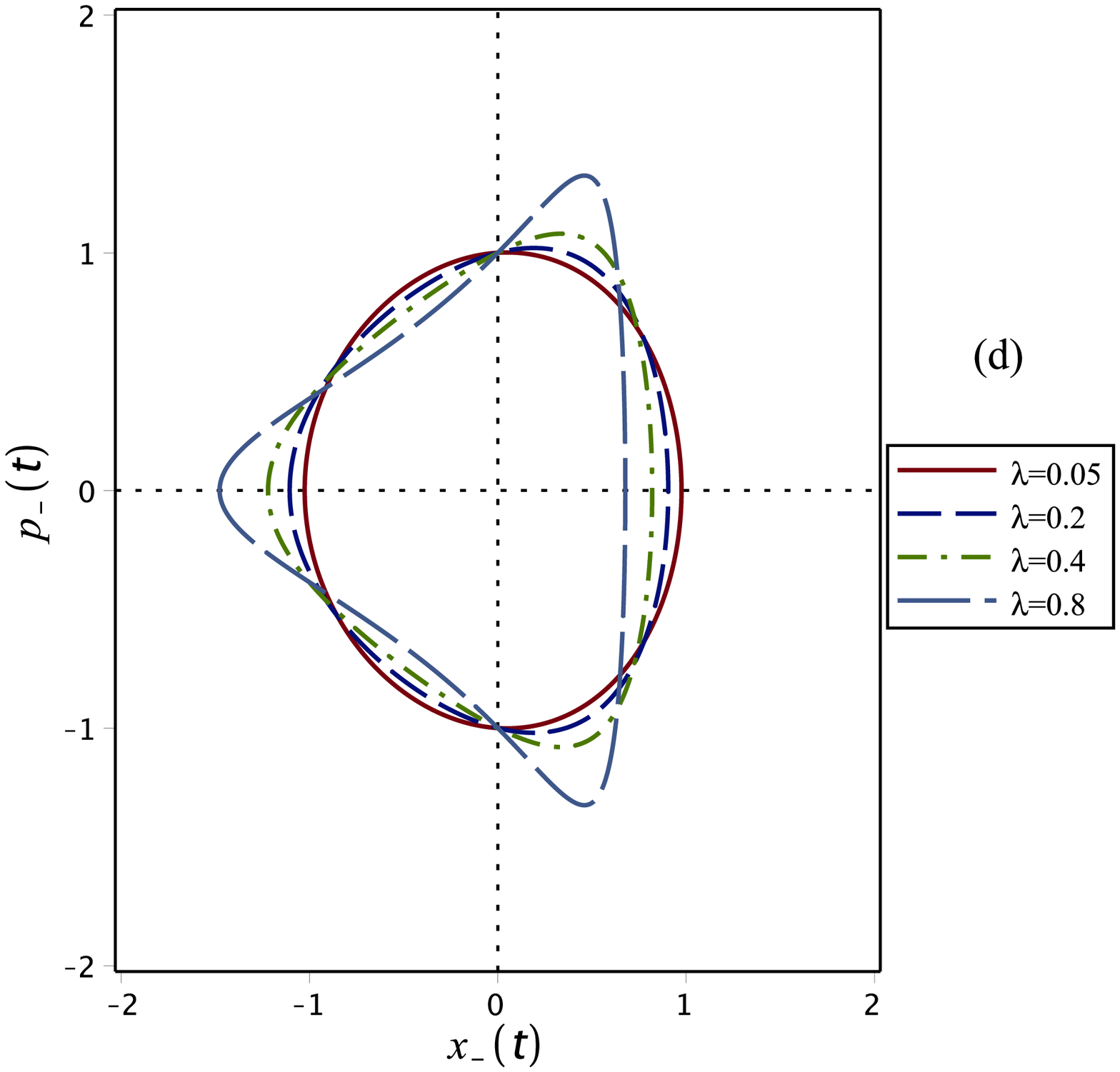}
\caption{\small 
{Shows (a) the Isochronous oscillator of $x_{+}(t)$ in (59), (b) the Phase trajectories of the Isochronous dynamical system in (60), $p_{+}(t)$ vs $x_{+}(t)$, (c) the Isochronous oscillator of $x_{-}(t)$ in (59),and (d) the Phase trajectories of the Isochronous dynamical system in (60), $p_{-}(t)$ vs $x_{-}(t)$.}}
\label{fig4}
\end{figure}
would imply that the PDM function is%
\begin{equation}
m\left( x\right) =-\frac{1}{4}\frac{\left( \lambda x-2\right) ^{2}}{\left(
\lambda x-1\right) ^{3}}
\end{equation}%

Using (50) and (39) one obtains%
\begin{equation}
q=\sqrt{Q(x)}%
x\iff x=x_{\pm}=\frac{A}{2}\cos \left( \omega t+\varphi \right) \left[
-\lambda A\cos \left( \omega t+\varphi \right) \pm \sqrt{\lambda
^{2}A^{2}\cos ^{2}\left( \omega t+\varphi \right) +4}\right] ,
\end{equation}%
which satisfies the corresponding dynamical equation,%
\begin{equation}
\ddot{x}-\frac{\lambda \left( \lambda x-4\right) }{2\left( \lambda
x-1\right) \left( \lambda x-2\right) }\dot{x}^{2}+\frac{2\left( \lambda
x-1\right) }{\lambda x-2}\,\omega ^{2}x=0,
\end{equation}%
and represents its exact isochronous nonlinear PDM-oscillator solution. We show in Figure 4 the corresponding two isochronous oscillators $x_{+}(t)$, 4(a), $x_{-}(t)$, 4(c), and the related phase trajectories for for the dynamical PDM equation (60).%
\begin{figure}[h!]  
\centering
\includegraphics[width=0.4\textwidth]{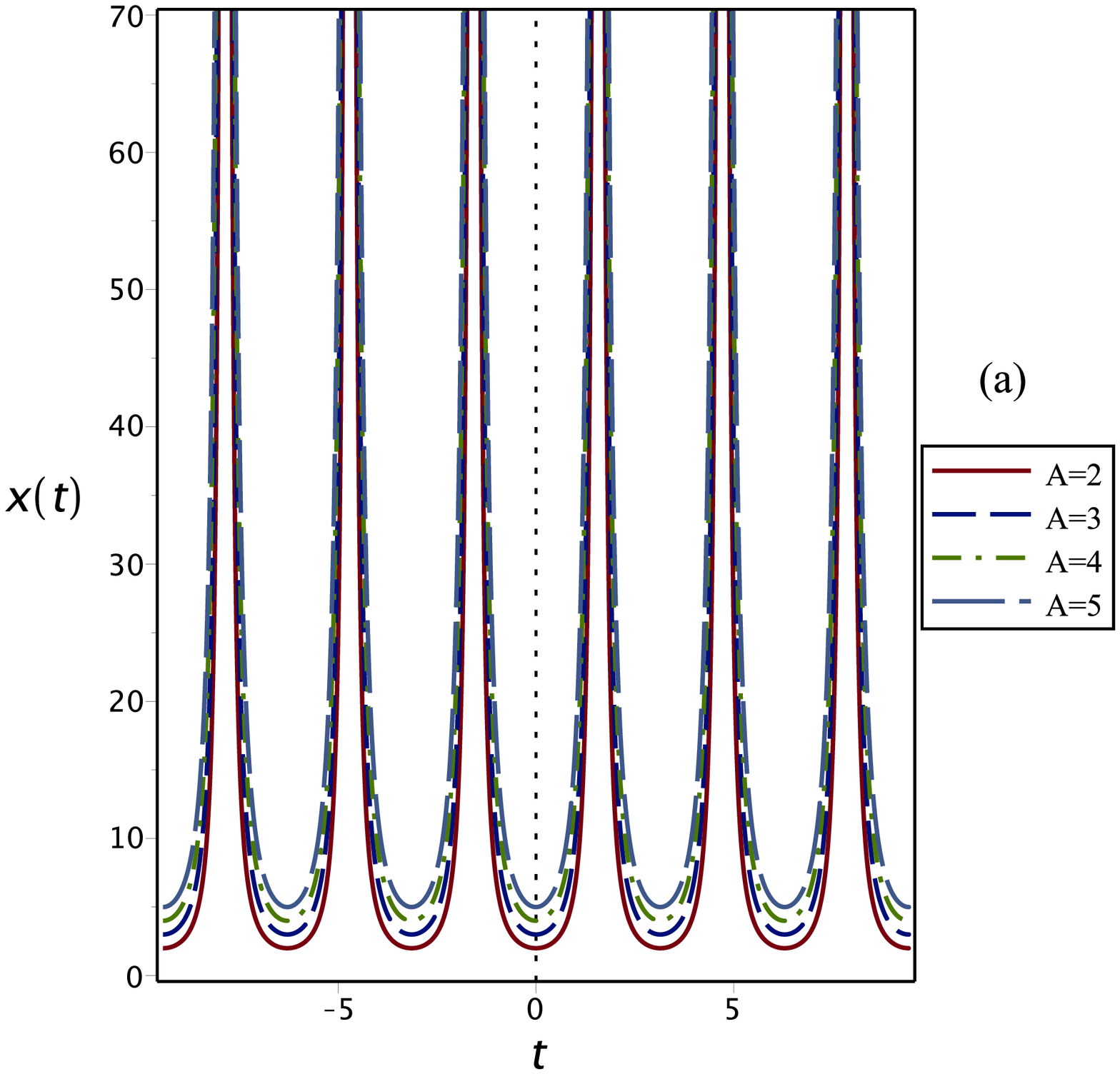}
\includegraphics[width=0.4\textwidth]{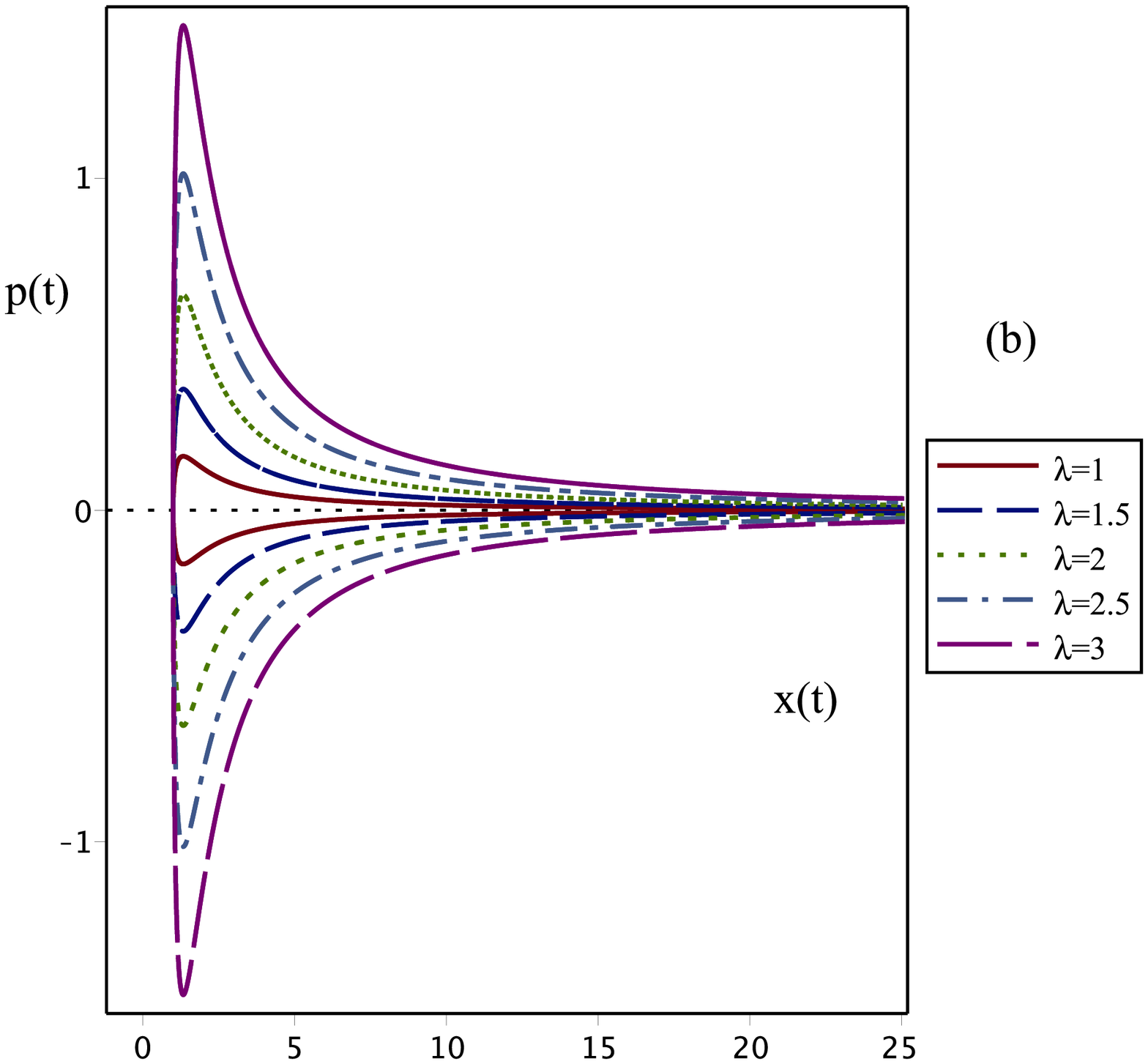}
\includegraphics[width=0.4\textwidth]{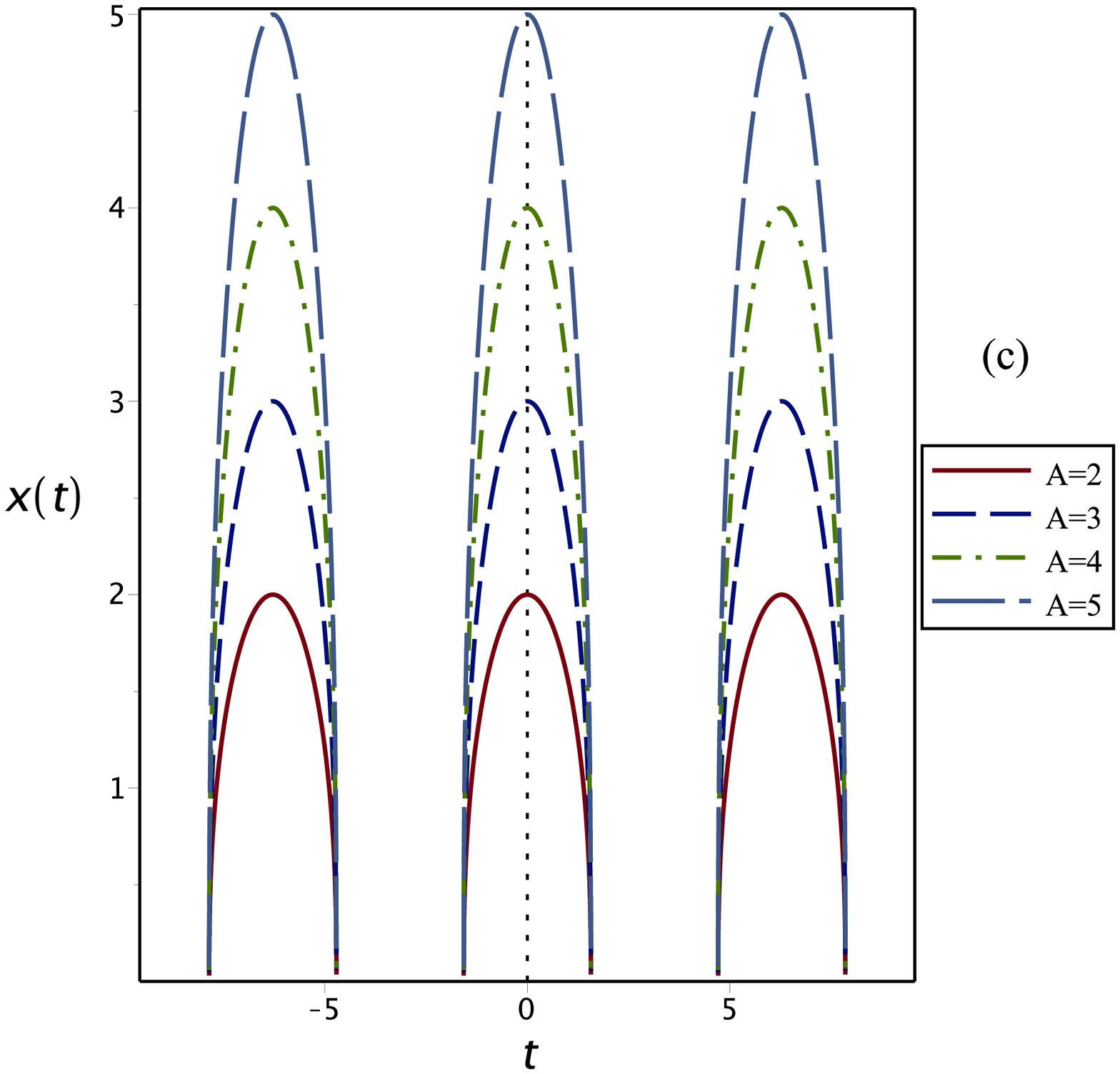}
\includegraphics[width=0.4\textwidth]{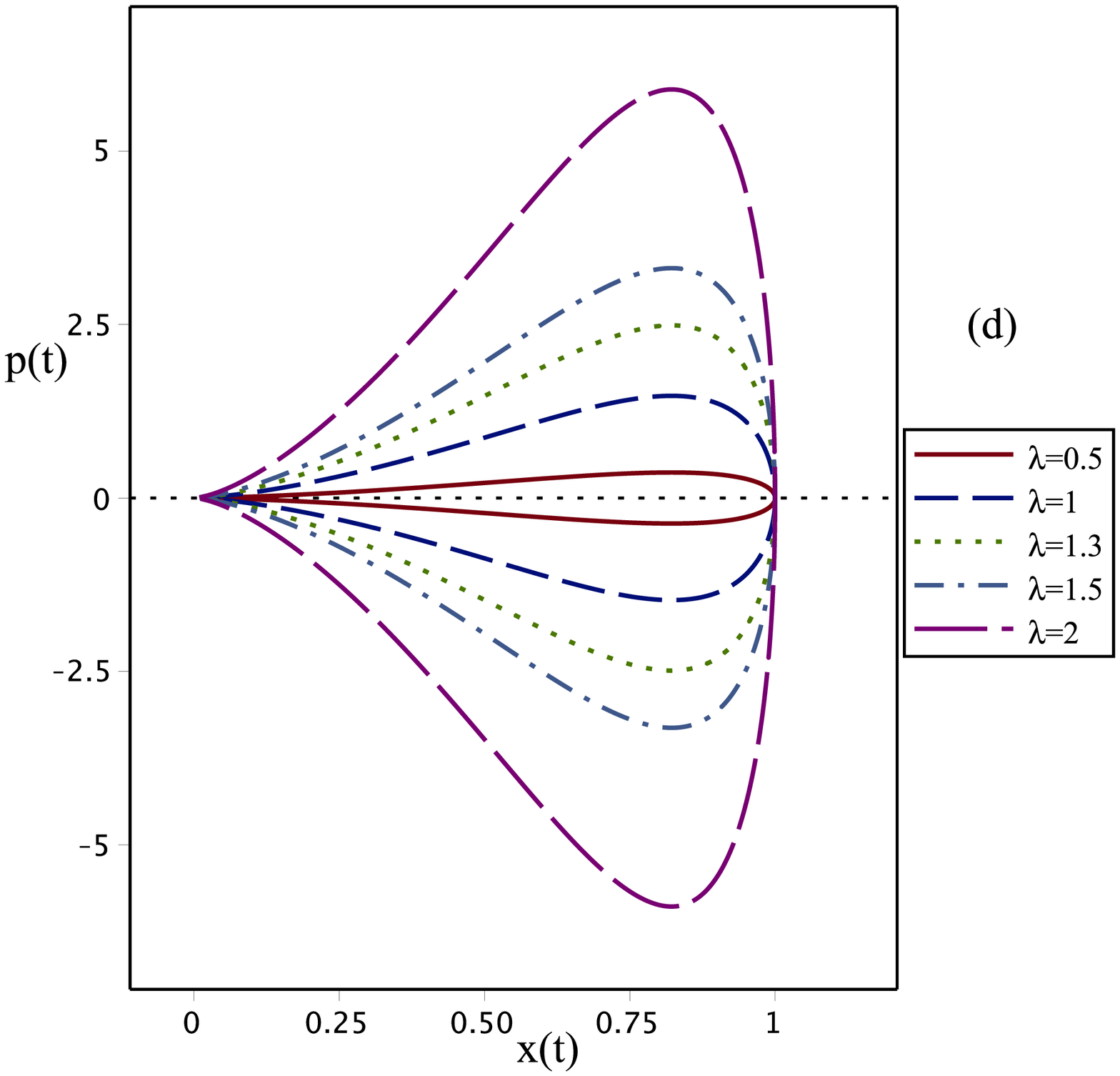}
\caption{\small 
{Shows (a) the Isochronous oscillator of $x(t)$ in (63) for $\upsilon=-3/2$, (b) the Phase trajectories of the isochronuous dynamical system in (64), $p(t)$ vs $x(t)$, for $\upsilon=-3/2$, (c) the Isochronous oscillator of $x(t)$ in (63) for $\upsilon=3/2$, and (d) the Phase trajectories of the isochronuous dynamical system in (64), $p(t)$ vs $x(t)$, for $\upsilon=3/2$.}}
\label{fig5}
\end{figure}
\subsubsection{A power-law type PDM: $m\left( x\right) \sim \,x^{2\upsilon }$}

A power-low type coordinate deformation%
\begin{equation}
\sqrt{Q\left( x\right) }=a\,x^{\upsilon }
\end{equation}%
would result the power-law type PDM function%
\begin{equation}
m\left( x\right) =a^{2}\left( \upsilon +1\right) ^{2}\,x^{2\upsilon }\text{ }%
;\text{  }\upsilon \neq -1.
\end{equation}%
Hence, using (50) and (39), the exact isochronous nonlinear PDM-oscillator solution would be%
\begin{equation}
q=A\cos \left( \omega t+\varphi \right) =a\,x^{\upsilon
+1}\Longleftrightarrow x=\left[ \frac{A}{a}\cos \left( \omega t+\varphi
\right) \right] ^{1/\left( \upsilon +1\right) },
\end{equation}%
that satisfies the dynamical equation, (47),%
\begin{equation}
\ddot{x}+\frac{\upsilon }{x}\dot{x}^{2}+\frac{1}{\upsilon +1}\,\omega ^{2}x=0%
\text{ };\text{  }\upsilon \neq -1.
\end{equation}%
\begin{figure}[h!]  
\centering
\includegraphics[width=0.4\textwidth]{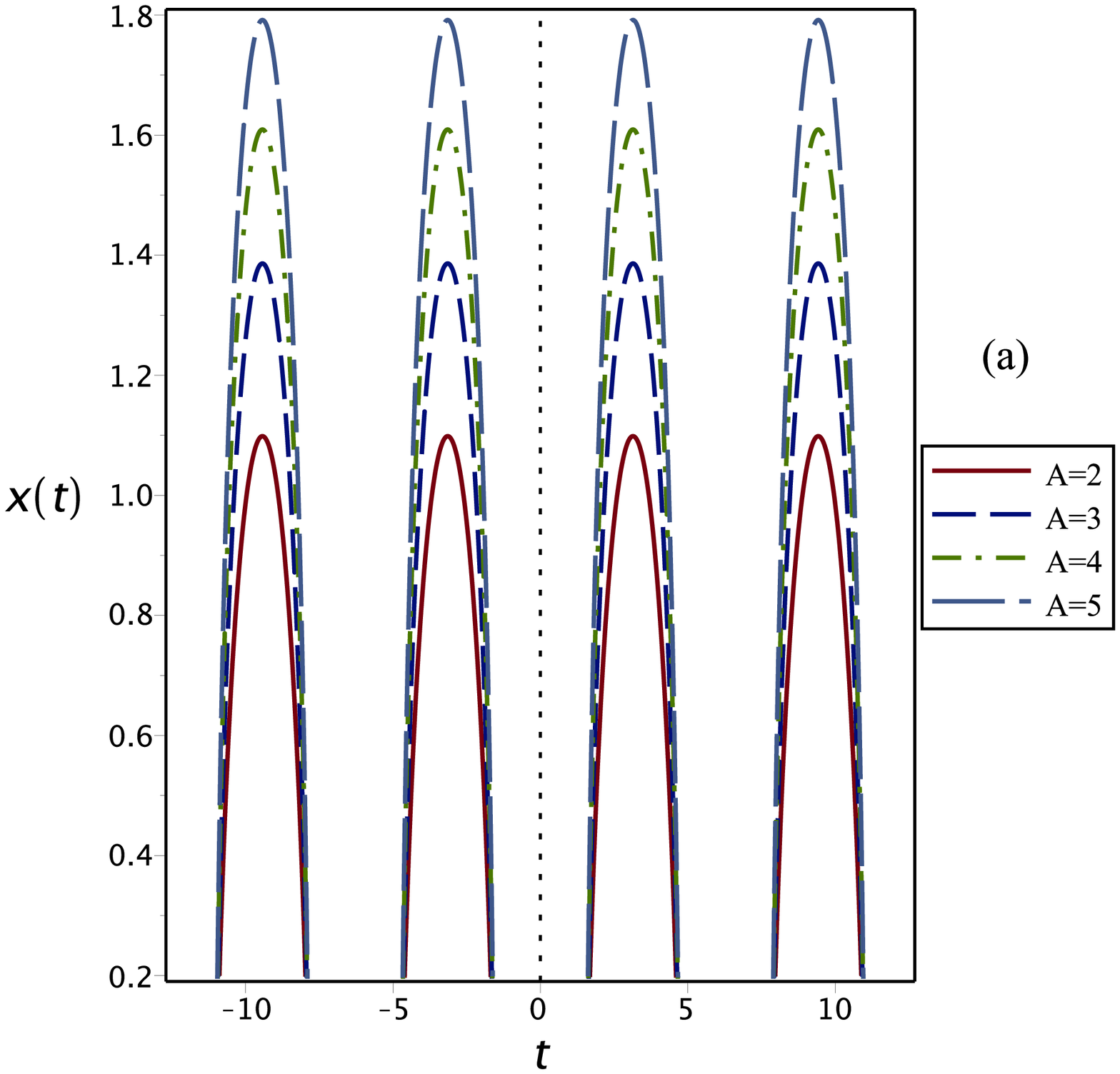}
\includegraphics[width=0.4\textwidth]{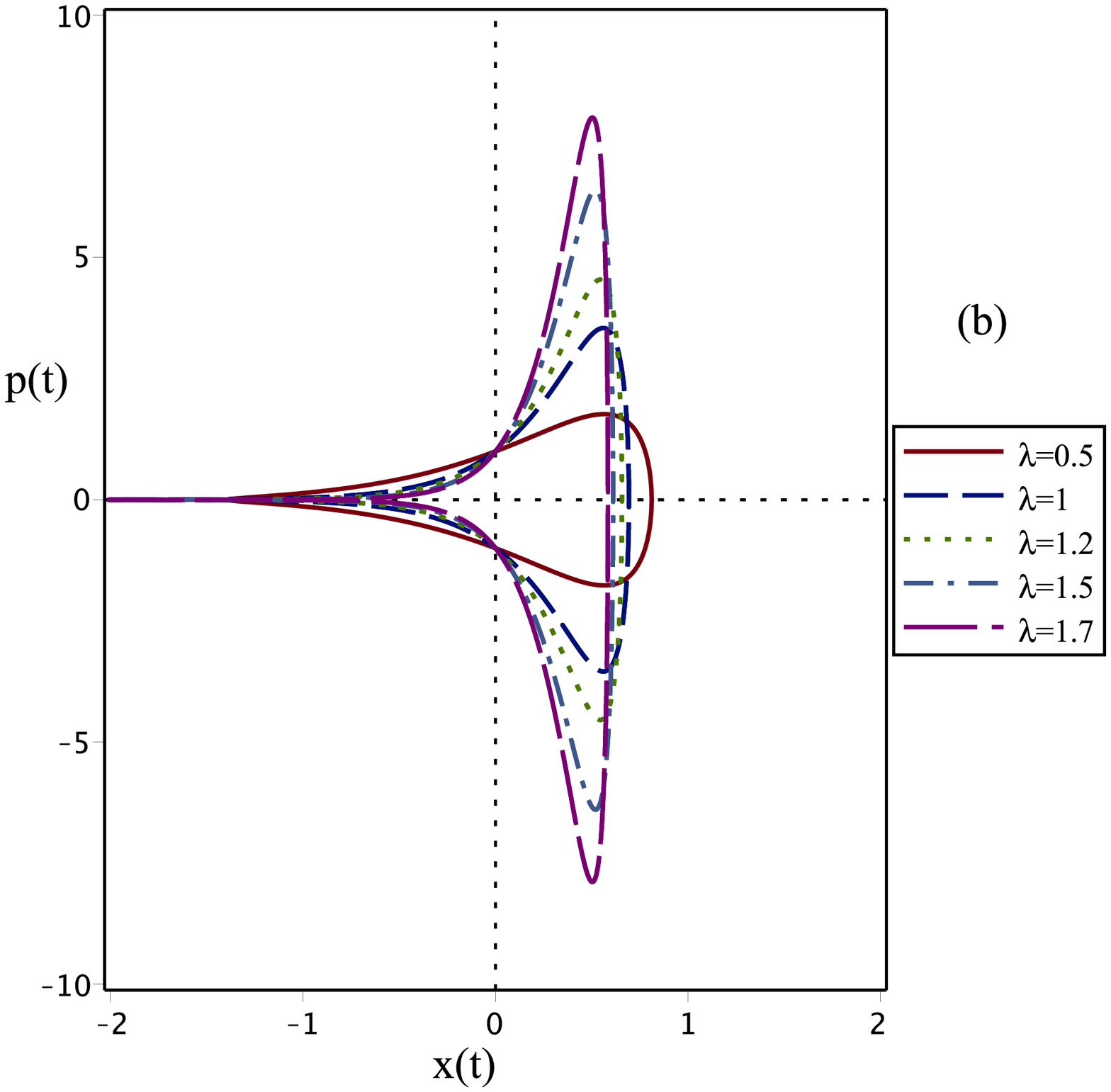}
\caption{\small 
{Shows (a) the Isochronous oscillator of $x(t)$ in (68), and (b) the Phase trajectories of the Isochronous dynamical system in (67).}}
\label{fig6}
\end{figure}
In Figure 5 we present the isochronous oscillator of $x(t)$ in (63) for $\upsilon=-3/2$ in Fig.5(a), and for $\upsilon=3/2$ in Fig.5(c). The Phase trajectories of the isochronous dynamical system (64) are plotted in Fig.5(b) for $\upsilon=-3/2$, and  Fig.5(d), for $\upsilon=3/2$.

\subsubsection{An exponential- type PDM: $m\left( x\right) =e^{2\lambda x}$}

An exponential-type PDM 
\begin{equation}
m\left( x\right) =e^{2\lambda x}
\end{equation}%
would imply, by (40), that the coordinate deformation is%
\begin{equation}
\sqrt{Q\left( x\right) }=\frac{e^{\lambda x}}{\lambda x}\left( 1-e^{-\lambda
x}\right); \lambda\neq 0 .
\end{equation}%
Which when substituted in the dynamical equation (51) yields%
\begin{equation}
\ddot{x}+\lambda \dot{x}^{2}+\,\frac{\omega ^{2}}{\lambda }\left(
1-e^{-\lambda x}\right) =0.
\end{equation}%
Using (50) and (39), one finds that it admits exact isochronous nonlinear
PDM-oscillator solution%
\begin{equation}
q=\,\frac{1}{\lambda }\left(
1-e^{\lambda x}\right) \Longleftrightarrow x=\frac{1}{\lambda }\ln \left(
1-\lambda A\cos \left( \omega t+\varphi \right) \right); \  0<\lambda<\frac{1}{A}.
\end{equation}%
In Figure 6 we present the isochronous oscillator of $x(t)$ of (68), and the Phase trajectories of the Isochronous dynamical system in (67).

\subsection{$n$-dimensional isochronous nonlinear PDM oscillators}

For the $n$-dimensional PDM-oscillators Lagrangian (47) case, we shall recollect that the Euler-Lagrange invariance falls short and incomplete. One has therefore to appeal to Newtonian invariance or $\dot{H}$-invariance and use the exact solution (50) of (49) to extract exact solutions for (47) (or equivalently (51).

\subsubsection{A coordinate deformations without singularity: $Q\left( r\right) =1/\left( 1+\lambda ^{2}r^{2}\right) $}

The coordinate deformations of the form%
\begin{equation}
\sqrt{Q\left( r\right) }=\sqrt{\frac{1}{1+\lambda ^{2}r^{2}}}\,;\,r=\sqrt{%
\sum\limits_{j=1}^{n}x_{_{j}}^{2}},
\end{equation}%
would result, by (40), the PDM function 
\begin{equation}
m\left( r\right) =\frac{1}{\left( 1+\lambda ^{2}r^{2}\right) ^{3}}.
\end{equation}%
This would allow us to write (39) as%
\begin{equation}
\mathbf{q} =\sqrt{Q\left( r\right) }%
\mathbf{r\Longleftrightarrow r=}\frac{\mathbf{q}}{\sqrt{1-\lambda ^{2}q^{2}}}%
\Longleftrightarrow x_{_{i}}=\frac{A_{_{i}}\cos \left( \omega t+\varphi
\right) }{\sqrt{1-\lambda ^{2}A^{2}\cos ^{2}\left( \omega t+\varphi \right) }%
};\,A=\sqrt{\sum\limits_{j=1}^{n}A_{_{j}}^{2}} ,  0<A<\frac{1}{\lambda}.
\end{equation}%
which satisfy our dynamical equations of (47)%
\begin{equation}
\ddot{x}_{_{i}}-\frac{6\lambda ^{2}}{1+\lambda ^{2}r^{2}}\left(
\sum\limits_{j=1}^{n}x_{_{j}}\dot{x}_{_{j}}\right) \,\dot{x}_{i}+\frac{%
3\lambda ^{2}}{1+\lambda ^{2}r^{2}}\left( \sum\limits_{j=1}^{n}\dot{x}%
_{_{j}}^{2}\right) \,x_{i}+\left( 1+\lambda ^{2}r^{2}\right) \omega
^{2}x_{i}=0,
\end{equation}%
and forms their exact $n$-dimensional isochronous nonlinear PDM-oscillators solutions, therefore.

\subsubsection{A power-law type PDM: $m\left( x\right) \sim \,r^{2\upsilon }$}

Consider a power-law type coordinate deformation%
\begin{equation}
\sqrt{Q\left( r\right) }=a\,r^{\upsilon },
\end{equation}%
which in turn implies a PDM function%
\begin{equation}
m\left( r\right) =a^{2}\left( \upsilon +1\right) ^{2}r^{2\upsilon }\text{ };%
\text{  }\upsilon \neq -1.
\end{equation}%
Consequently, with $\mathbf{q=A}\cos \left( \omega t+\varphi \right) $, equation (39) yields%
\begin{equation}
\mathbf{q=}a\,r^{\upsilon }\mathbf{r\Longleftrightarrow r=}\left( \frac{%
q^{-\upsilon }}{a}\right) ^{1/\left( \upsilon +1\right) }\mathbf{q}%
\Longleftrightarrow x_{_{i}}=A_{_{i}}\cos \left( \omega t+\varphi \right)
\,\left( \frac{\left[ A\cos \left( \omega t+\varphi \right) \right]
^{-\upsilon }}{a}\right) ^{1/\left( \upsilon +1\right) },
\end{equation}%
as the exact $n$-dimensional isochronous nonlinear PDM-oscillators solutions for the dynamical equations (47) 
\begin{equation}
\ddot{x}_{_{i}}+\frac{2\upsilon }{r^{2}}\left( \sum\limits_{j=1}^{n}x_{_{j}}%
\dot{x}_{_{j}}\right) \,\dot{x}_{i}-\frac{\upsilon }{r^{2}}\left(
\sum\limits_{j=1}^{n}\dot{x}_{_{j}}^{2}\right) \,x_{i}+\frac{\omega ^{2}}{%
\upsilon +1}x_{i}=0\,;\,r^{2}=\sum\limits_{j=1}^{n}x_{_{j}}^{2}.
\end{equation}

In the sample of illustrative example discussed above, we notice that there are no constraints on the frequencies of the nonlinear PDM oscillators considered. Such frequencies are clearly amplitude-independent and are isochronic. Therefore, all our examples are isochronous nonlinear PDM oscillators.

\section{Concluding Remarks}

In this work, we have considered the $n$-dimensional PDM-Lagrangians in their standard form ( i.e., the difference between kinetic and potential energies). However, in order to make our study comprehensive and self-contained, we have recollected and elaborated on the solubility and linearizability of the non-standard PDM Mathews-Lakshmanan nonlinear oscillators (6). The generalization of such
nonlinear ML-oscillators (6) to any PDM, $m\left( r\right) $, settings is also discussed and reported in section II. Yet, we have asserted that the
position-dependent deformation of time (manifested by the nonlocal point transformations in (22) or (23)) renders such PDM nonlinear oscillators non-isochronous (i.e., their frequencies become amplitude-dependent). We have also reported the standard Lagrangian form for the ML- PDM $m(x)=1/(1+\lambda x^{2})$ along with the standard PDM potential force field (19) and the corresponding isochronous dynamical equation (20).  The comparison between the standard and non-standard PDM potentials clearly indicates that the non-standard ML-PDM potential may support oscillatory motion over a narrow region in space, whereas the standard one supports oscillatory motion over the full space (documented in Figures 1(a), 2(a), and (2(b)).  

To preserve isochronicity of the PDM nonlinear oscillators in $n$-dimensions, we had to return back to the standard Lagrangians form to obtain a set of interesting isochronous PDM nonlinear oscillators. In so doing, we have shown and emphasized (in section III) that negative the gradient of the PDM potential force field (i.e., the force vector associated with PDM settings) is no longer given by the time derivative of the canonical momentum, $\mathbf{p}\left( \mathbf{r}%
\right) =m\left( r\right) \mathbf{\dot{r}}$, but it is rather given in terms of the pseudo-momentum, $\mathbf{\pi }\left( r\right) =\sqrt{m\left(
r\right) }\mathbf{\dot{r}}$ \cite{5,7} (or the Noether momentum as in \cite{6}). That is,
\begin{equation*}
-\mathbf{\nabla }V\left( \mathbf{q}\left( \mathbf{r}\right) \right) =\mathbf{%
F}=\sqrt{m\left( r\right) }\frac{d}{dt}\left( \sqrt{m\left( r\right) }%
\mathbf{\dot{r}}\right) ,
\end{equation*}%
In the same section, moreover, we have shown that the connection between constant mass settings and PDM settings is feasible through some point transformation, where the time is kept as is (i.e., no position-dependent deformation of time). Hereby, the Euler-Lagrange invariance is shown satisfactory for $n=1$ but unsatisfactory/incomplete for $n\geq 2$. Hence, alternative invariances are sought through either the total Newtonian vector form or the total time derivative of the PDM Hamiltonian. Consequently, in addition to "\textit{Newtonian invariance}" of Mustafa \cite{42}, we have introduced yet another type of invariance to be called, hereinafter, "$\dot{H}$-\textit{invariance}" (where $\dot{H}=dH/dt$). Moreover, such invariances go alongside with the fact that the total energy is a conserved quantity (documented in (43), (44), and (45)) and is a constant of motion, therefore. This result allowed us to use, in section IV and V, the well known exact solutions (50) of the linear oscillator (49), along with our point transformation (38), to obtain exact solutions for a set of $n$-dimensional isochronous nonlinear PDM oscillators. This is documented in the illustrative examples of section V, where a set of one-dimensional
and a set of $n$-dimensional isochronous nonlinear PDM oscillators are reported.

On the isochronicity and linearizability sides of the quadratic  Li\'{e}nard-type nonlinear differential equation (9), it is unavoidably obligatory to make the connection between our point transformation proposal (38) and the Lie point symmetries approach of Tiwari et al. [2] on the isochronicity condition and linearizability of (9). Obviously, one finds that our $q(x)$ is $h(x)$ of Tiwari and our $G(x)$ and $F(x)$ are $g(x)$ and $f(x)$ of Tiwari, respectively. Therefore, if one uses our $F(x)=m'(x)/2m(x)=h''(x)/h'(x)$ and $G(x)=\omega^{2}\sqrt{Q(x)/m(x)}x=\omega^{2}h(x)/h'(x)$  then the results in sections IV and V of Tiwari et al. [2] are retrieved and recovered. For example, Tiwari et al.'s generalized Morse oscillator (their equation (43)) is our dynamical system in (67) with a PDM $m(x)=e^{2\lambda x}$. Their dynamical equation (49) corresponds to a PDM $m(x)=1/(1+\lambda x)^{4}$  and  $Q(x)=1/(1+\lambda x)^{2}$ and so on.  Consequently, with $F(x)=m'(x)/2m(x)$ and $G(x)=\omega^{2}\sqrt{Q(x)/m(x)}x$, our methodical proposal provides a general format for the standard PDM Lagrangians that results isochronuous quadratic  Li\'{e}nard-type nonlinear differential equations (9), which are indeed linearizabile under our point transformation recipe (38).  To the best of our knowledge, our general results and/or methodical proposal have never been reported elsewhere in the literature.

\end{document}